\documentclass[12pt,draftclsnofoot,onecolumn,twoside]{IEEEtran}
\usepackage[mathscr]{eucal}
\usepackage{ifpdf}
\usepackage{cite}
\usepackage{graphicx}
\usepackage[cmex10]{amsmath}
\usepackage{amssymb}
\usepackage{algorithmic}
\usepackage{units}
\usepackage{setspace}
\usepackage{algorithm}
\usepackage{array}
\usepackage[tight,footnotesize]{subfigure}
\usepackage{amsthm}
\usepackage{multirow}
\usepackage{enumerate}
\usepackage{color}
\usepackage{mathtools}
\usepackage{subfigure}

\newcounter{assume}

\newtheorem{theorem}{Theorem}
\newtheorem{lemma}{Lemma}

\newtheorem{assumption}[assume]{Assumption}
\newtheorem{remark}{Remark}

\newtheorem{definition}{Definition}

\newcommand{\resl}[1]{}
\newcommand{\tosl}[1]{}
\newcommand{\cosl}[1]{}
\newcommand{\insl}[1]{#1}

\begin{document}
\title{Channel Selection for Network-assisted D2D Communication via No-Regret Bandit 
Learning with Calibrated Forecasting}
\author{
\IEEEauthorblockN{Setareh Maghsudi and S\l awomir Sta\'{n}czak, \textit{Senior Member, IEEE}\\}

\thanks{Parts of the material in this paper were presented at the IEEE
  Workshop on Signal Processing Advances in Wireless Communications,
  Darmstadt, June, 2013. The work was supported by the German Research
  Foundation (DFG) under grant STA 864/3-2. The authors are with the
  Fachgebiet f\"ur Informationstheorie und theoretische
  Informationstechnik, Technische Universit\"at Berlin, and with the
  Fraunhofer Institute for Telecommunications Heinrich Hertz
  Institute, Berlin, Germany (e-mail: setareh.maghsudi@tu-berlin.de,
  slawomir.stanczak@hhi.fraunhofer.de).}%
}
\maketitle
\begin{abstract}
We consider the distributed channel selection problem in the context of 
device-to-device (D2D) communication as an underlay to a cellular network. 
Underlaid D2D users communicate directly by utilizing the cellular spectrum 
but their decisions are not governed by any centralized controller. Selfish 
D2D users that compete for access to the resources construct a distributed 
system, where the transmission performance depends on channel availability 
and quality. This information, however, is difficult to acquire. Moreover, 
the adverse effects of D2D users on cellular transmissions should be minimized. 
In order to overcome these limitations, we propose a network-assisted distributed 
channel selection approach in which D2D users are only allowed to use vacant 
cellular channels. This scenario is modeled as a multi-player multi-armed 
bandit game with side information, for which a distributed algorithmic solution 
is proposed. The solution is a combination of no-regret learning and calibrated 
forecasting, and can be applied to a broad class of multi-player stochastic 
learning problems, in addition to the formulated channel selection problem. 
Analytically, it is established that this approach not only yields vanishing 
regret (in comparison to the global optimal solution), but also guarantees 
that the empirical joint frequencies of the game converge to the set of 
correlated equilibria. 
\end{abstract}


\section{Introduction}\label{sec:Introduction}
%
\subsection{Related Works}\label{subsec:Related}
D2D communication underlying cellular networks enables wireless devices to communicate 
directly instead of through an access point or a base station (BS), provided that such 
transmissions do not disturb prioritized cellular transmissions \cite{Fodor12}. This 
concept has been proposed to (i) boost the overall network efficiency expressed in terms 
of radio resources, and (ii) to provide users with more reliable services at lower costs \cite{Lei12}.\\ 
Similar to other wireless networking scenarios, spectrum resource management is a key 
component of D2D communication systems, and may be studied within two different frameworks. 
The first framework, adapted by \cite{Feng13}, \cite{Han12} and \cite{Wang12} among many 
others, considers D2D and cellular systems as the two parts of a single entity with resources 
allocated by some BS, which is assumed to be in possession of global channel and network 
knowledge. The second framework, in contrast, considers a hierarchical relationship, where 
cellular and D2D systems are regarded as primary and secondary systems, respectively, and 
the resource allocation for the D2D system is performed in a distributed manner. This viewpoint 
can be found some literatures including \cite{Xu122}, \cite{Xu13}, \cite{Wang13} and \cite{Yaacoub12}.\\ 
Since conventional pilot signals cannot be used for estimation of D2D channels, the assumption 
of precise D2D channel information availability at some BS is not realistic. As a result, we 
argue in favor of the second framework described before, and we assume that D2D users establish 
a secondary \textit{distributed} network that is allowed to use \textit{vacant} spectrum resources 
of cellular network, thereby causing \textit{no} interference to primary users.
In this context, a vast majority of schemes, including those proposed by papers mentioned above, 
have some game-theoretical basis. Most of the game-theoretical models, however, require that the 
players know at least their own utility function. Moreover, statistical knowledge on channel gains 
and/or traffic model should be available. If not, they either require heavy information exchange 
among users (buyer-seller market models \cite{Maghsudi12b}, cooperative game models \cite{Khan11}), 
and/or a coordinator (auctions \cite{Sodagari11}). In addition, since game-theoretical framework 
(cooperative or non-cooperative) requires all game parties to be known by each other, provision 
of required information is extremely costly. In order to address these shortcomings, one approach 
is to incorporate learning theory, as performed in the context of cognitive radio networks. Some 
works, for instance \cite{Guo04}, \cite{Fang13} and \cite{Song07}, consider single-agent learning 
scenarios, while others study opportunistic spectrum access in multi-agent learning setting. In 
such setting, agents have access to strictly limited or even no information, and the objective is 
to satisfy some optimality condition. In what follows, we discuss some of these works in more details.   

In \cite{XuA12}, opportunistic spectrum access is formulated as a multi-agent learning game. 
In this work, it is assumed that upon availability, each channel pays the same reward to all 
users. This assumption, however, is strictly restrictive as it neglects channel qualities. 
On the other hand, if a channel is selected by multiple users, orthogonal spectrum access 
is applied, and therefore interference is neglected. In \cite{XuO13} and \cite{XuS13}, authors 
consider the interference minimization game for partially overlapping channels. In these works, 
it is assumed that interference emerge only between neighboring users, and the proposed learning 
approaches are based on graphical games. In addition, in the three works mentioned above, the 
designed game is proven to be an exact potential game so that a pure-strategy Nash equilibrium 
exists. It can be thus concluded that the generalization of proposed approaches as well as 
convergence analyses is not straightforward. In \cite{Liang12}, two approaches are proposed to 
achieve Nash equilibrium in a multi-player cognitive environment. System verification, however, 
is only based on numerical approaches. Other examples are \cite{KalathilO12}, \cite{KalathilD12} 
and \cite{Kalathil14}. In these works, channel qualities are taken into account; nonetheless, 
it is assumed that in case of collision, no reward is paid to colliding users. Thus, interference 
is again neglected. Moreover, in the proposed algorithms, learning and channel selection are two 
independent procedures; while the former follows multi-armed bandit scenario, the latter is 
formulated as bipartite graph matching. This decoupling yields unnecessary complexity, and it 
is also not clear whether the final solution is stable or not, which is the main concern of 
equilibrium. The works \cite{LiuS10}, \cite{LiuN10} and \cite{LiuM13} propose various selection 
schemes to achieve logarithmic regret as well as fairness among users. However, equilibrium 
analysis is absent.

\subsection{Our Contribution}\label{subsec:Contribution}
In this paper, we study a multi-player adaptive decision making problem, where selfish players 
learn the optimal action from successive interactions with a dynamic environment, and finally 
settle at some equilibrium point. This problem appears in many wireless networking scenarios, 
with a particular instance being the channel selection in a distributed D2D communication system 
integrated into a centralized cellular network. In our setting, each D2D user is selfish and 
aims at optimizing its throughput performance, while being allowed to use \textit{vacant} cellular 
channels. We model this problem as a multi-armed bandit game among multiple learning agents 
that are provided with \textit{no} prior information about channel quality and availability. 
We propose a channel selection strategy that consists of two main blocks, namely calibrated 
forecasting (\cite{Mannor10}, \cite{Kakade08}, \cite{Foster96}) and no-regret bandit learning 
(\cite{Yang02}, \cite{Bianchi06}, \cite{singh00}, \cite{Chapman11}). Whereas calibrated forecasting 
is utilized to predict the joint action profile of selfish rational players, no-regret learning 
builds a trust-worthy estimate of the reward generating processes of arms. We show that our proposed 
model and selection strategy can be applied to both noise-limited (orthogonal channel access) 
and interference-limited (non-orthogonal channel access) transmission models.
We prove that the gap between the 
average utility achieved by our approach and that of the optimal fixed strategy converges to 
zero as the game horizon tends to infinity. Moreover, by using our strategy, the empirical 
joint frequencies of play converge to the set of correlated equilibria.\\
As discussed in Section \ref{subsec:Related}, the spectrum access problem using learning 
theory has been under extensive study in recent years. Nevertheless, our work differs from 
previous studies in many aspects, as listed briefly in the following. 
\begin{itemize}
\item Some works such as \cite{Guo04} and \cite{Fang13} analyze single-agent learning problem. 
In some others such as \cite{KalathilO12} and \cite{KalathilD12}, although multi-agent problem 
is formulated, no explicit equilibrium analysis is performed. We, however, propose an algorithmic 
solution for multi-agent learning and show that by applying our approach the empirical joint 
frequencies of the game converge to the set of correlated equilibria. As any Nash equilibrium 
belongs to the set of correlated equilibria, our solution is more general in comparison to approaches 
that converge to a pure-strategy Nash equilibrium, for example those proposed in \cite{XuA12}, 
\cite{XuO13}, \cite{XuS13} and \cite{Liang12}.
\item The proposed multi-player learning approach can be applied to solve a wide range of 
resource allocation problem, including radio resource management, routing, scheduling, object 
tracking and so on. This is due to the fact that our convergence analysis \textit{does not} 
depend on utility or cost function. In contrast, References \cite{XuA12}, \cite{XuO13} and 
\cite{XuS13} require the game be an \textit{exact potential} game for an equilibrium to be 
achieved by proposed approaches, and hence the applicability of these approaches is strictly 
restricted.   
\item In our problem setting, both noise-limited and interference-limited transmission models are 
studied, and
do not impose any limitation on the interference pattern. This is in contrast with \cite{KalathilD12}, 
\cite{XuO13} and \cite{XuS13}, where the interference is either completely neglected or is limited 
to neighboring users. This is important since depending on channel matrices, channel allocation 
based on interference avoidance might be suboptimal.
\item In our work, channels (or generally, actions) differ for different users. More precisely, 
variations in both channel availability and quality is taken into account. This stands in contrast 
to \cite{XuA12}, where the average gain of each specific channel is assumed to be equal for all 
users (deterministic), and only availability is considered to be stochastic. 
\end{itemize}

\subsection{Paper Structure}\label{subsec:Organization}
The paper is organized as follows. Section \ref{sec:System} includes system model and problem 
formulation. In Section \ref{sec:Problem}, we present basic elements of bandit games, and model 
the formulated problem as a multi-player multi-armed bandit game. Section \ref{sec:Calibration} 
briefly reviews calibrated forecasting. In Section \ref{sec:BanditGame}, we propose our channel 
selection strategy. Section \ref{sec:Numerical} includes numerical results, while Section \ref{sec:conclusion} 
concludes the paper.

\section{System Model and Problem Formulation}\label{sec:System}
\subsection{System Model}\label{subsec:SystemModel}
We study a distributed D2D communication system as an underlay to a cellular network. 
The D2D system consists of $K$ device pairs referred to as D2D users, denoted by either 
just $k$ or the pair $(k,k')$. The single-cell wireless network is provided with $M$ 
licensed orthogonal channels. In such network structure, cellular users\footnote{Cellular 
users are those users who communicate via base stations.}~and D2D users are regarded 
as primary and secondary, respectively. As a result, a channel is available to D2D 
users only if it is not occupied by any cellular user. D2D users have \textit{neither} 
channel (quality and availability) \textit{nor} network (traffic) knowledge. We assume 
that the BS observes the transmission channels of all D2D and cellular users. D2D users 
do not exchange information. However, there exists a control channel through which the 
BS broadcasts some signals referred to as side information, which is heard by all D2D 
users. This assumption is justified by the physical characteristics of the radio propagation 
medium. Note that the control channel is occupied only until convergence, and therefore 
the overhead remains low. Throughout the paper, $h_{uv,m}(t)$ stands for the coefficient 
of channel $m$ (including Rayleigh fading and path loss) between nodes $u$ and $v$ at time 
$t$. The variance of zero-mean additive white Gaussian noise (AWGN) is denoted by $N_{0}$.
\subsection{Transmission Model}\label{subsec:TransmissionModel}
The transmission structure of D2D users is described in the following. At each transmission round, 
D2D user selects a channel to sense (\textit{selection} phase). For simplicity, we assume that 
sensing is perfect. Afterwards \textit{transmission} phase begins. Primary duration of this phase 
is denoted by $T_{r}$; however, as we see shortly, the useful transmission time of D2D user depends 
on channel availability and the applied multiple access technique. After the transmission phase 
\textit{announcement} phase begins, in which the BS broadcasts D2D indices (IDs) along with indices 
of their selected channels. Consequently, all D2D users know which users have transmitted in each 
channel.\footnote{Later we see that this side information helps D2D users to converge to an efficient 
stable point.}~This phase is followed by \textit{learning} phase. In the learning phase, every D2D 
user exploits its gathered data, including its achieved throughput and also the received broadcast 
message, to learn the environment as well as strategies of other players.

Since all D2D users are allowed to select among $M$ (probably available) channels, collision 
might occur. We consider the following multiple access protocols. 
\begin{itemize}
\item Orthogonal multiple access (noise-limited region): If multiple D2D users select a common 
channel, carrier sense multiple access (CSMA) is implemented to address collision issues \cite{ZhaoA07}, 
\cite{XuA12}. Since interference is avoided, transmissions are corrupted only by AWGN. Therefore 
the throughput of some D2D user $k$ transmitting at some channel $m$ yields
\begin{equation}
\label{eq:TDMA}
R_{m,k}(t,k^{-})=\frac{\tau_{m,k}(k^{-})}{T_{r}}\log\left (1+\frac{P\left |h_{kk',m}(t)\right|^{2}}
{N_{0}} \right)I_{m}, 
\end{equation}
where $k^{-}$ denotes the set of channels selected by all D2D users except for $k$, and 
$\tau_{m,k}(k^{-})$ is a random variable that stands for the useful transmission time of 
user $k$ through channel $m$. The probability density function (pdf) of $\tau_{m,k}(k^{-})$ 
depends on the exact applied CSMA scheme, and is not calculated here since it impacts neither 
applicability nor analysis. An example of such calculations can be found in \cite{XuA12}. 
$I_{m}$ is a Bernoulli random variable with parameter $\theta_{m}$ that indicates whether 
channel $m$ is occupied by some cellular user or not.
\item Non-orthogonal multiple access (interference-limited region): If multiple D2D users select 
a common channel, they all transmit together, which results in interference. In this case, the 
throughput of D2D user $k$ is given by
\begin{equation}
\label{eq:CDMA}
R_{m,k}(t,k^{-})=\log\left (1+\frac{P\left |h_{kk',m}(t)\right|^{2}}{P\sum_{l=1}^{L(k^{-})}\left |h_{lk',m}(t) \right|^{2}+N_{0}}\right)I_{m},
\end{equation}
where $L(k^{-})$ denotes the number of D2D users that share channel $m$ with user $k$.
\end{itemize}
\subsection{Problem Formulation}\label{subsec:ProblemFormulation}
Let $m_{k,t}$ and $k^{-}_{t}$ respectively denote the selected channel of D2D user 
$k$ and the set of channels selected by all D2D users except for $k$, both at time 
$t$, yielding $R_{m_{k,t},k}(t,k_{t}^{-})$. Ideally, at every time 
$t$, D2D user $k$ selects the optimal channel in the sense of maximum throughput, 
thereby maximizing its accumulated throughput. Therefore, its \textit{ultimate} goal 
can be formulated as 
\begin{equation}
\label{eq:FirstGoal}
\underset{m_{k,t} \in \left \{1,...,M \right \}}{\textup{maximize}}~\sum_{t=1}^{T}R_{m_{k,t},k}(t,k^{-}_{t}),
\end{equation}
with $T$ being the total transmission time. However, since D2D users have no prior 
information, solving (\ref{eq:FirstGoal}) can be notoriously difficult or even impossible. 
Consequently, we argue in favor of another strategy where each D2D user pursues a less 
ambitious goal: Minimize its regret, which is the difference between the throughput that 
\textit{could have been achieved} by selecting the optimal channel (if it were known), and 
that of the actual selected channel. We formulate this problem as follows. Let $m^{\circ}_{k,t}
\coloneqq\arg\max_{m \in \left \{1,..,M \right\}}R_{m,k}(t,k_{t}^{-})$ be the optimal channel 
that yields a throughput equal to $R^{\circ}_{k}(t,k^{-}_{t})\coloneqq R_{m^{\circ}_{k,t},k}
(t,k^{-}_{t})$. Since $m^{\circ}_{k,t}$ is not known, D2D user $k$ attempts to choose 
a channel whose reward is asymptotically as large as $m^{\circ}_{k,t}$. This can be 
formalized as
\begin{equation}
\label{eq:SecondGoal}
\lim_{T \to \infty}\frac{1}{T}\sum_{t=1}^{T} \left (R_{m_{k,t},k}(t,k^{-}_{t})-R^{\circ}_{k}(t,k^{-}_{t}) \right)=0.
\end{equation}
Let $f_{m,k}(k^{-})=\textup{E}_{t}\left [R_{m,k}(t,k^{-})\right]<+\infty$ for $m \in 
\left \{1,...,M \right \}$, where $\textup{E}_{t}$ denotes the expected value over time. 
Furthermore, let $m^{*}_{k,t} \coloneqq\arg\max_{m \in \left \{1,..,M \right\}}f_{m,k}
(k_{t}^{-})$ that results in $f^{*}_{k}(k^{-}_{t})\coloneqq f_{m^{*}_{k,t},k}(k^{-}_{t})$. 
In \cite{Yang02}, it is shown that (\ref{eq:SecondGoal}) is equivalent to
\begin{equation}
\label{eq:ThirdGoal}
\lim_{T\to \infty}\frac{1}{T}\sum_{t=1}^{T}\left(f_{m_{k,t},k}(k^{-}_{t})-f^{*}_{k}(k^{-}_{t})\right)=0,
\end{equation}
provided that $R_{m,k}(t,k^{-})$ is bounded above and away from zero. In this paper, 
we assume that each D2D user $k$ aims at satisfying (\ref{eq:ThirdGoal}) as the 
performance metric.

%
\section{Bandit-Theoretical Model of Resource Allocation Problem}\label{sec:Problem}
\subsection{Single-Player and Multi-Player Multi-Armed Bandit}\label{subsec:MP_MAB}
Single-player multi-armed bandit game (SP-MAB, hereafter) is a class of sequential 
decision making problems with limited information. In a SP-MAB setting, a player has 
access to a finite set of actions (arms, interchangeably). Upon being pulled by the 
player at time $t$, each arm, say arm $m \in \left \{1,...,M \right\}$, generates a 
random reward $R_{m}(t) \in \mathbb{R}^{+}$. The player only observes the reward of 
the played arm, and not those of other arms. We denote the continuous mean reward 
associated with $m$ by $f_{m}$, $m \in \left \{1,...,M \right \}$. That is, 
$f_{m}=\textup{E}_{t} \left [R_{m}(t)\right]$. We denote the optimal arm and its 
associated mean reward by $m^{*}$ and $f^{*}$ respectively, where we define 
$f^{*}\insl{\coloneqq}\max_{m \in \left \{1,..,M \right\}} f_{m}$. The player needs 
to decide which action to take at successive rounds in a way that asymptotically the 
accumulated reward achieved by the played arms is not much less than that of the 
optimal arm. Obviously, this problem is an instance of the well-known exploitation-exploration 
dilemma, in which a balance should be found between exploiting the arms that have 
exhibited good performance in the past (control), and exploring arms that might 
perform well in the future (learning).
In multi-player multi-armed bandits (MP-MAB, hereafter), this formulation remains unchanged, 
and players still face exploration-exploitation dilemma. The problem, though, becomes more 
challenging, since in multi-player settings with reward sharing, the rewards achieved by any 
player do not only depend on the arms pulled by this player, but also on actions of other 
players. Hence, for player $k$ that pulls arm $m$, the mean reward is denoted by
$f_{m,k}(k^{-})=\textup{E}_{t}\left [R_{m,k}(t,k^{-})\right]$, where $k^{-}$ denotes the 
joint action profile of all players other than $k$ (that is, its opponents), which has 
$M^{K-1}$ realizations. In other words, the reward achieved by player $k$ at time $t$ yields 
$R_{m_{k,t},k}(t,k^{-}_{t})$, where $m_{k,t}$ denotes the selected action, and $k^{-}_{t}$ 
is the realization of the joint action profile of opponents, both at time $t$. At each trial 
$t$, for player $k$, we denote the optimal arm and its associated mean reward by $m^{*}_{k,t}$ 
and $f^{*}_{k}(k^{-}_{t})$ respectively, where we have $f^{*}_{k}(k^{-}_{t})\insl{\coloneqq}
\max_{m \in \left \{1,..,M \right\}} f_{m,k}(k^{-}_{t})$ and $m^{*}_{k,t}\coloneqq \arg 
\max_{m \in \left \{1,..,M \right\}} f_{m,k}(k^{-}_{t})$. We assume that $f_{m,k}(k^{-})$ and 
$f^{*}_{k}(k^{-})$ obey the following assumption \cite{Yang02}.
\begin{assumption}
\label{as:expected_reward}
$ \forall ~k \in \left \{1,...,K \right\}$, $m \in \left \{1,...,M \right\}$, $k^{-} \in 
\bigotimes_{k'=1,k'\neq k}^{K}\left \{1,...,M \right \}$ ($\bigotimes$ denotes the Cartesian 
product) and $t>0$,
\begin{enumerate}[a)]
\item $f_{m,k}(k^{-})\in [0,A]$ for some $A>0$,
\item $B=\sup_{m}\sup_{k^{-}}\left(f^{*}_{k}(k^{-})-f_{m,k}(k^{-})\right)<\infty $,
\item $\textup{E}(f^{*}_{k}(k^{-}_{1}))> 0$.
\end{enumerate}
\end{assumption}
%
The last part of the assumption implies that the expected optimal reward is positive at least for the first 
round of the game, which is used later to avoid division by zero later. We also assume that the achieved 
rewards of any particular player are revealed to that player only, while actions of players can be observed 
by their opponents.

At the $T$-th play, the collection of personal achieved rewards and observed actions 
up to time $T$, are available to each player. The accumulated mean reward of player $k$ up 
to time $T$ is $\sum_{t=1}^{T}f_{m_{k,t},k}(k^{-}_{t})$, while $\sum_{t=1}^{T}f^{*}_{k}(k^{-}_{t})$ 
is the optimal total reward of player $k$,  which could have been achieved by pulling 
arm $m^{*}_{k,t}$ for all trials\resl{$t$} up to $T$. Since the $k$-th player would 
attain the best performance if it selected at every trial the optimal arm, it is reasonable 
to evaluate any selection strategy $\kappa$ used by player $k$ based on the following performance 
metric of interest given by \cite{Yang02}
\begin{equation}
\label{eq:consistency_metric}
S_{\kappa,T}=\frac{\sum_{t=1}^{T}f_{m_{k,t},k}(k^{-}_{t})}{\sum_{t=1}^{T}f^{*}_{k}(k^{-}_{t})} \leq 1\,,
\end{equation}
where $\sum_{t=1}^{T}f^{*}_{k}(k^{-}_{t})>0$ by Assumption \ref{as:expected_reward}. Clearly, the closer 
$S_{\kappa,T}$ to 1, the better the selection strategy. Asymptotically as $T$ tends to infinity, the most 
desired property is \textit{strong consistency}, defined below.
\begin{definition}[\cite{Yang02}]
\label{def:strong_consistency}
A selection strategy $\kappa$ is strongly consistent if $S_{\kappa,T}\to 1$ as $T \to \infty$.
\end{definition}
\begin{remark}[\cite{Yang02}]
\label{re:equivalency}
If $\frac{1}{T}\sum_{t=1}^{T}f^{*}_{k}(k^{-}_{t})$ is bounded above and away from $0$ with 
probability $1$, then $S_{\kappa,T}\to 1$ almost surely is equivalent to $\lim_{T\to \infty}\frac{1}{T}\sum_{t=1}^{T}\left(f_{m_{k,t},k}(k^{-}_{t})-f^{*}_{k}(k^{-}_{t})\right)=0$; 
referring to "$f_{m_{k,t},k}(k^{-}_{t})-f^{*}_{k}(k^{-}_{t})$" as "regret" at time $t$, this 
implies that strong-consistency is equivalent to achieving per-round vanishing (zero-average) 
regret.
\end{remark}
From the game-theoretic point of view, for each player $k \in \left \{1,...,K \right\}$, an 
MP-MAB can be seen as a game with two agents: the first agent is player $k$ itself, and the 
second agent is the set of all other $K-1$ players whose joint action profile affects the 
rewards of player $k$. Since the reward of any player $k$ depends on the decisions of other 
players, a key idea of the proposed approach is to enable each user to \textit{forecast} the 
future actions of its opponents based on public knowledge. In Section \ref{sec:Calibration}, 
we discuss how reliable forecasting can be performed and how players should proceed using 
this side information. 
\subsection{Modeling the Channel Selection Problem as Bandit Game}\label{subsec:BanditModel}
By comparing our system model (Section \ref{sec:System}) with MP-MAB (Section \ref{subsec:MP_MAB}), 
we observe that distributed channel selection problem is in great harmony with MP-MAB settings. 
Therefore, we model this problem as an MP-MAB game, in which each D2D user is modeled as a player, 
while frequency channels are regarded as arms, and choosing a channel is pulling an arm. Clearly, 
the instantaneous reward achieved by any player, which is its attained throughput,\footnote{Throughput 
is considered as an exemplary reward function, and it can be substituted by any other utility or 
cost function.}~depends on the selected channel of the player itself and also on those of other 
players, with the throughput given by (\ref{eq:TDMA}) under orthogonal multiple access strategies 
and (\ref{eq:CDMA}) when non-orthogonal strategies are used. By Remark \ref{re:equivalency}, the 
goal of D2D users, which is (\ref{eq:ThirdGoal}), is equivalent to strong-consistency.

\section{Calibration and Construction of a Calibrated Forecaster}\label{sec:Calibration}
At each time $t$, any D2D user $k$ is aware of actions of other $K-1$ players up to time $t-1$. 
Using this knowledge, it attempts to predict the joint action of others at time $t$, to minimize 
the harm of opponents on its reward, by taking the best-response to the predicted joint action 
profile. For prediction, we use calibrated forecasting, for a reason that is stated formally in 
Theorem \ref{th:CorrConvergence}. This theorem states that by using calibrated forecaster, we 
ensure the possibility of achieving an equilibrium point. In what follows, we describe calibrated 
forecasting briefly. 
\subsection{Calibration}\label{subsec:calibration}
Following \cite{Mannor10}, consider a random experiment with a finite set of outcomes $\mathfrak{D}$ 
of cardinality $\textup{D}$, and let $\delta_{d_{t}}$ stand for the Dirac probability distribution on 
some outcome $d$ at time $t$. The set of probability distributions over $\mathfrak{D}$ is denoted by 
$\mathfrak{P}=\Delta (\mathfrak{D})$, $\mathfrak{P}\subseteq \mathbb{R}^{\textup{D}}$. Equip $\mathfrak{P}$ 
with some norm $\left \| \cdot  \right \|$. At time $t$, forecaster outputs a probability distribution 
$\mathbf{P}_{t}$ over the set of outcomes.
\begin{definition}[\cite{Mannor10}]
\label{de:Calibration}
A forecaster is said to be calibrated if $\forall~\epsilon>0$ and
$\forall~\mathbf{p}\in \mathfrak{P}$, almost surely, 
\begin{equation}
\label{eq:calibration}
\lim_{T\to \infty }\left \| \frac{1}{T} \sum_{t=1}^{T}\mathbb{I}_{\left \{ \left \| \mathbf{P}_{t}-\mathbf{p} \right \|\leq \epsilon  \right \}}\left (\mathbf{P}_{t}-\delta_{d_{t}} \right)\right \|=0.
\end{equation}
\end{definition}

A relaxed notion of calibration is $\epsilon$-calibration. Given $\epsilon>0$, 
an $\epsilon$-calibrated forecaster considers some finite covering of $\mathfrak{P}$ by 
$N_{\epsilon}$ balls of radius $\epsilon$. Denoting the centers of these balls 
by $\mathbf{p}_{1},...,\mathbf{p}_{N_{\epsilon }}$, the forecaster selects only 
forecasts $\mathbf{P}_{t}\in \left \{\mathbf{p}_{1},...,\mathbf{p}_{N_{\epsilon}} \right \}$. 
Using this, $\epsilon$-calibration is defined as follows.
\begin{definition}[\cite{Mannor10}]
\label{de:Epsilon_Calibration}
Define $Q_{t}$ to be the index in $\left \{1,...,N_{\epsilon} \right \}$ such that 
$\mathbf{P}_{t}=\mathbf{p}_{Q_{t}}$. A forecaster is said to be $\epsilon$-calibrated 
if almost surely,
\begin{equation}
\label{eq:Epsilon_Calibration}
\limsup_{T\to \infty }\sum_{q=1}^{N_{\epsilon }}\left \| \frac{1}{T} \sum_{t=1}^{T}\mathbb{I}_{\left \{Q_{t} =q \right \}}\left ( \mathbf{p}_{q}-\delta _{d_{t}} \right )\right \|\leq \epsilon.
\end{equation}
\end{definition}
Note that none of the two definitions makes any assumption on the nature of the random experiment 
whose outcome is being predicted. The following result can be found in \cite{Foster96}, \cite{Kakade08}, 
and \cite{Bianchi06}.
\begin{theorem}
\label{th:CorrConvergence}
Consider a game with $K$ players provided with $M$ actions. Let $\mathfrak{C}$ stand for the set of 
correlated equilibria, and define the joint empirical frequencies of play as
\begin{equation}
\label{eq:empiricalFreq}
\hat{\pi}_{T}(\textbf{m})=\frac{1}{T}\sum_{t=1}^{T}\mathbb{I}_{\left \{ \textbf{M}_{t}=\textbf{m} \right \}},~~\textbf{m}=(m_{1},..,m_{K})\in \bigotimes_{k=1}^{K}\left \{ 1,..,M \right \},
\end{equation}
where $\textbf{M}_{t}$ denotes the joint action profile of players at time $t$, and $\bigotimes$ is 
the Cartesian product. Now, assume that each player plays by best responding to a calibrated forecast 
of the opponents joint action profile in a sequence of plays; that is, for each player $k$ we have
\begin{equation}
\label{eq:BestResponse}
m_{k,t}=\underset{m \in \left \{1,...,M \right \}}{\arg\max}~~\sum_{d=1}^{D}p_{d,k,t}f_{m,k}(d),
\end{equation} 
where $\mathbf{P}_{k,t}=(p_{1,k,t},...,p_{D,k,t})$ stands for the output of its forecaster, 
which is a probability distribution over $D=M^{K-1}$ possible joint action profiles of its 
opponents. Accordingly, each $d$ represents a realization of the joint action profile of 
opponents of player $k$, i.e. $k^{-}$. Then the distance $\inf _{\pi \in \mathfrak{C}}\sum_{\textbf{m}}\left | \hat{\pi}_{T}(\textbf{m})-\pi(\textbf{m})\right |$ between the empirical joint distribution 
of plays and the set of correlated equilibria converges to $0$ almost surely as $T \to \infty$.
\end{theorem}
\subsection{Construction of a Calibrated Forecaster}\label{subsec:construction}
For constructing a calibrated forecaster, an approach is to use \textit{doubling-trick} 
\cite{Mannor10}. In the first step, an $\epsilon$-calibrated forecaster is constructed 
for some $\epsilon>0$. Then, the time is divided into periods of increasing length, 
and the procedure of $\epsilon$-calibration is repeated as a sub-routine over periods, 
where $\epsilon$ decreases gradually at each period (that is, $N_{\epsilon}$-grid 
becomes finer), until it reaches zero. In Algorithm \ref{alg:calibrated_forecaster}, 
we review this procedure. The proof of calibration follows from the \textit{Blackwell's 
approachability} theorem. See \cite{Mannor10} for details and the proof of calibration.
\begin{algorithm}
\caption{A Calibrated Forecaster \cite{Mannor10}}
\label{alg:calibrated_forecaster}
\small
\begin{algorithmic}[1]
\STATE Define an increasing sequence of integers, $T_{r}=2^r$, $r=1,2,...$. Each member $T_{r}$ of this sequence 
denotes the length of period $r$, i.e. the number of trials included in it.
\STATE For each period $r$, let $\epsilon_{r}=2^{-r/(\textup{D}+1)}$.
\STATE Define a game, where the first player is the forecaster with the action 
set $I=\left \{1,...,N_{\epsilon_{r}} \right\}$ and the second player is the 
nature with action set $J=\mathfrak{D}$, $\left |\mathfrak{D} \right |=\textup{D}$. 
The first player is in fact some D2D user $k$, and the second player is the set 
of all other $K-1$ D2D users, i.e. its opponents. Moreover, $D=M^{K-1}$. Also, 
any outcome $d$ is the realization of a joint action profile of $K-1$ players, 
that is $k^{-}$.
\STATE Define the vector-valued regret of the first player as $u\left \{q,d \right \}=\left ( \mathbf{0},...,\mathbf{0},\mathbf{p}_{q}-\delta_{d},\mathbf{0},...,\mathbf{0} \right )$ 
for each $q \in \left \{ 1,...,N_{\epsilon_{r}} \right \}$, $d \in \mathfrak{D}$.
\STATE Define the target set $\mathfrak{F}$ as follows:
\begin{itemize}
\item Write ($\textup{D}N_{\epsilon_{r}}$)-dimensional vectors of $\mathbb{R} ^{\textup{D}N_{\epsilon_{r}}}$ as 
$N_{\epsilon_{r}}$-dimensional vectors with components in $\mathbb{R} ^{\textup{D}}$, i.e. $\mathbf{X}=\left (\mathbf{x_{1}},...,\mathbf{x_{N_{\epsilon_{r}}}} \right)$, where $\mathbf{x_{l}} \in \mathbb{R}^{\textup{D}}$ for all $l\in \left \{1,...,N_{\epsilon_{r}} \right \}$.
\item $\mathfrak{F}$ is a subset of the $\epsilon_{r}$-ball around $\left (\mathbf{0},...,\mathbf{0} \right)$ 
for the calibration norm $\left \| \cdot  \right \|$, which is a closed convex set.
\end{itemize}
(The forecaster is $\epsilon_{r}$-calibrated when this set is approachable by the 
regret vector (see \cite{Mannor10} for details)).
\STATE Define the sequence of the vector-valued regrets up to time $T$ ($1\leq T \leq T_{r}$) as
%
\begin{equation}
\mathbf{u}_{T}=\frac{1}{T}\sum_{t=1}^{T}u(Q_{t},d_{t})=
\frac{1}{T} \left (\sum_{t=1}^{T}\mathbb{I}_{\left \{Q_{t}=1 \right \}}(\mathbf{p}_{1}-\delta_{t}),..., 
\sum_{t=1}^{T}\mathbb{I}_{\left \{Q_{t}=N_{\epsilon_{r}} \right \}}(\mathbf{p}_{N_{\epsilon_{r}}}-\delta_{t})\right ).
\end{equation}
Now, (\ref{eq:Epsilon_Calibration}) (condition of $\epsilon_{r}$-calibration) can be restated as the converges 
of $\mathbf{u}_{T_{r}}$ to the set $\mathfrak{F}$ almost surely. In the following, $\mathbf{u}^{(r)} \coloneqq 
\mathbf{u}_{T_{r}}$ denotes the final regret of period $r$. 
\REPEAT
       \FOR {$t=1 \to T_{r}$}
            \IF{($r=1 \wedge  t=1$)}
                \STATE Pick up an action $Q_{t}$ from $I$ according to uniform distribution over the action set, 
                       i.e. let $\psi_{1}=(\frac{1}{N_{\epsilon_{r}}},...,\frac{1}{N_{\epsilon_{r}}})$. Note that 
                       $\psi_{t}$ is the mixed strategy at time $t$, while $\psi^{(r)} \coloneqq \psi_{T_{r}}$ 
                       denotes the final mixed strategy of period $r$.
            \ELSIF{($r>1 \wedge  t=1$)}
                 \STATE Pick up an action $Q_{t}$ from $I$ according to a probability distribution in a small 
                        neighborhood of $\psi^{(r-1)}$ (localization of search).
            \ELSE 
                  \STATE Pick up an action $Q_{t}$ from $I$ at random according to a distribution $\psi_{t}=\left \{\psi_{t,1},...,\psi_{t,N_{\epsilon_{r}}}                                \right\}$ on $\left \{1,...,N_{\epsilon_{r}} \right \}$ such that $\forall d\in \mathfrak{D}$ 
                  \begin{equation}
                  \label{eq:prob_density_psi}
                  \left (\mathbf{u}_{t-1}-\Pi_{\mathfrak{F}}(\mathbf{u}_{t-1}) \right )\cdot \left (u(\psi_{t},d)-\Pi_{\mathfrak{F}}(\mathbf{u}_{t-1}) \right                         )\leq 0,
                  \end{equation}
                  where $\Pi_{\mathfrak{F}}$ denotes the projection in $l^{2}$-norm onto $\mathfrak{F}$ and $\cdot$ 
                  denotes the inner product in $\mathbb{R}^{\textup{D}N_{\epsilon_{r}}}$. See \cite{Mannor10} and 
                  \cite{Freund99} for details.
            \ENDIF
        \ENDFOR
        \STATE Calculate the final regret of the current period, $\mathbf{u}^{(r)}=\mathbf{u}_{T_{r}}$. Also, let 
               $\psi^{(r)}=\psi_{T_{r}}$.
        \IF {$\mathbf{u}^{(r)} > \epsilon_{r}$,}
            \STATE  
            \begin{itemize}
            \item Let $r=1$ and $t=1$.
            \end{itemize}
        \ELSE
            \STATE
             \begin{itemize}           
             \item Let $r=r+1$ and $t=1$. 
             \end{itemize}
        \ENDIF
\UNTIL{convergence ($r$ has increased enough so that $\epsilon_{r}\approx 0$)}      
\end{algorithmic}
\end{algorithm}
%
\begin{theorem}[\cite{Mannor10}]
\label{th:CalibAlgorithm}
The forecasting procedure (Algorithm \ref{alg:calibrated_forecaster}) is calibrated. That is, it 
satisfies (\ref{eq:calibration}) (and with it (\ref{eq:Epsilon_Calibration})).
\end{theorem}

\section{Bandit Game}\label{sec:BanditGame}
As it is clear from (\ref{eq:TDMA}) and (\ref{eq:CDMA}), the throughput performance depends on two 
factors: 1) channel quality and availability, which is not affected by D2D users, and 2) number of 
D2D users transmitting in each channel, which is determined by the actions of users. Initially, none 
of these factors is known and their impact on the reward should be learned over time. For a D2D user 
$k$, the \textit{true} mean reward function of a channel $m \in \left \{ 1,...,M \right \}$ can be 
modeled as $f_{m,k}(k^{-})+\varepsilon_{m,k}$, where $\varepsilon_{m,k}$ denotes a random error with 
zero mean and finite variance \cite{Yang02}, independent over time, channels and users. Regardless 
of the type of regression analysis, here we make the following assumption.
\begin{assumption}
\label{as:consistent_reg}
The regression process is strongly consistent in $L_{\infty}$ norm for each $f_{m,k}(k^{-})$; 
that is, $\|\hat{f}_{m,k,t}(k^{-})-f_{m,k}(k^{-})\|_{\infty}\to 0$, for all $1\leq m\leq M$, 
$1\leq k\leq K$ and $k^{-} \in \bigotimes_{k'=1,k'\neq k}^{K}\left \{1,...,M \right \}$, 
almost surely as $t \to \infty$, where $\hat{f}_{m,k,t}(k^{-})$ denotes the regression estimate 
of $f_{m,k}(k^{-})$ at the $t$-th trial.
\end{assumption}
In Section \ref{subsec:BanditModel}, we modelled the channel selection as a bandit game. In what follows, 
we describe our proposed strategy to solve this game and investigate its convergence characteristics.
\subsection{Selection Strategy}\label{subsec:selection}
The game horizon is first divided into periods $r=1,2,...$ of increasing length 
$T'_{r}$. Moreover, we define another sequence $Z_{r}$ for $r=1,2,...$, so that 
$T'_{r}$ and $Z_{r}$ satisfy the following assumption.
\begin{assumption}
\label{as:bandit_game}
\begin{enumerate}[a)]
$\left\{T'_{r}\right\}_{r=1,2,...}$ and $\left\{Z_{r}\right\}_{r=1,2,...}$ are selected so that
\item $\left \{ \left \lceil T'_{r}Z_{r} \right \rceil \right \}_{r=1,2,...}$ is an increasing sequence of integers,
\item $\lim_{R \to \infty} \sum_{r=1}^{R}\left \lceil T'_{r}Z_{r} \right \rceil \rightarrow \infty$,
\item $\lim_{R \to \infty} \frac{\sum_{r=1}^{R}\left \lceil T'_{r}Z_{r} \right \rceil}{\sum_{r=1}^{R}T'_{r}}=0$.
\end{enumerate}
\end{assumption}
At each period $r$, $\left \lceil T'_{r}\cdot Z_{r} \right \rceil$ randomly-selected trials are 
devoted to exploration, and the rest of the trials are used for exploitation, in the following 
manner.  
\begin{itemize}
\item Exploitation: In an exploitation trial, say $i$, every player $k$ first receives 
a probability distribution $\mathbf{P}_{i}$ over all possible joint action profiles of 
other $K-1$ players, which is the output of its forecasting procedure. Based on this 
information, and by using the estimated mean reward functions, it selects the action 
with the highest estimated expected reward; that is, it acts with the best-response 
to the predicted joint action profile of its opponents.
\item Exploration: In an exploration trial, say $j$, with probability $\gamma \ll 1$, 
again best-response is played (see above), while with probability $1-\gamma$, an action 
is selected uniformly at random. 
\end{itemize}
In all trials, after selection, the player's estimation of the reward process of the 
selected action is upgraded based on the achieved reward. Moreover, actions of other 
players are observed (here by hearing the broadcast message). This observation is used 
by the forecaster, as described in Algorithm \ref{alg:calibrated_forecaster}. The 
entire procedure is summarized in Algorithm \ref{alg:Selection}.

\begin{figure}[ht]
\centering
\subfigure[Exemplary exploration-exploitation trade-off.]{%
\includegraphics[width=0.45\textwidth]{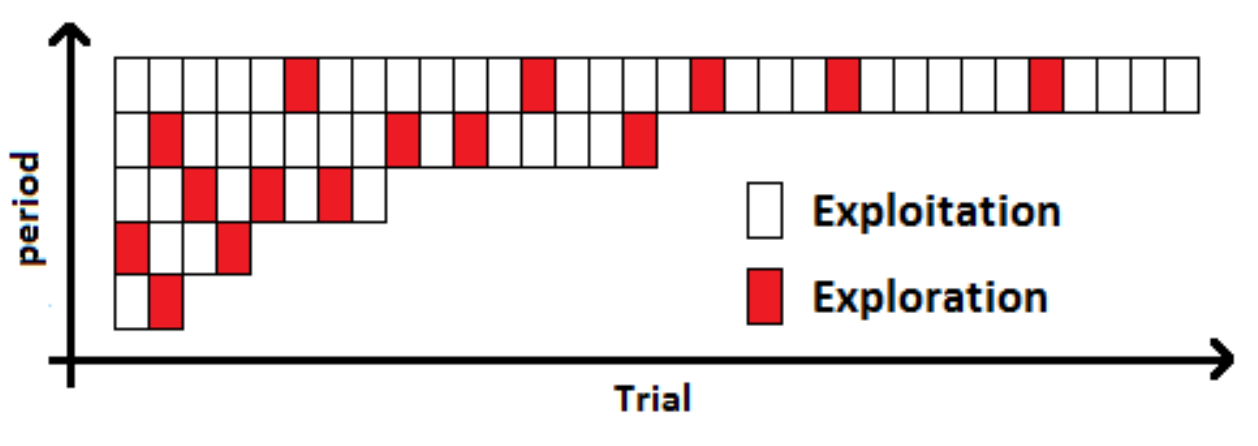}
\label{EE}}
\quad
\subfigure[Selection strategy flowchart (Algorithm \ref{alg:Selection}).]{%
\includegraphics[width=0.35\textwidth]{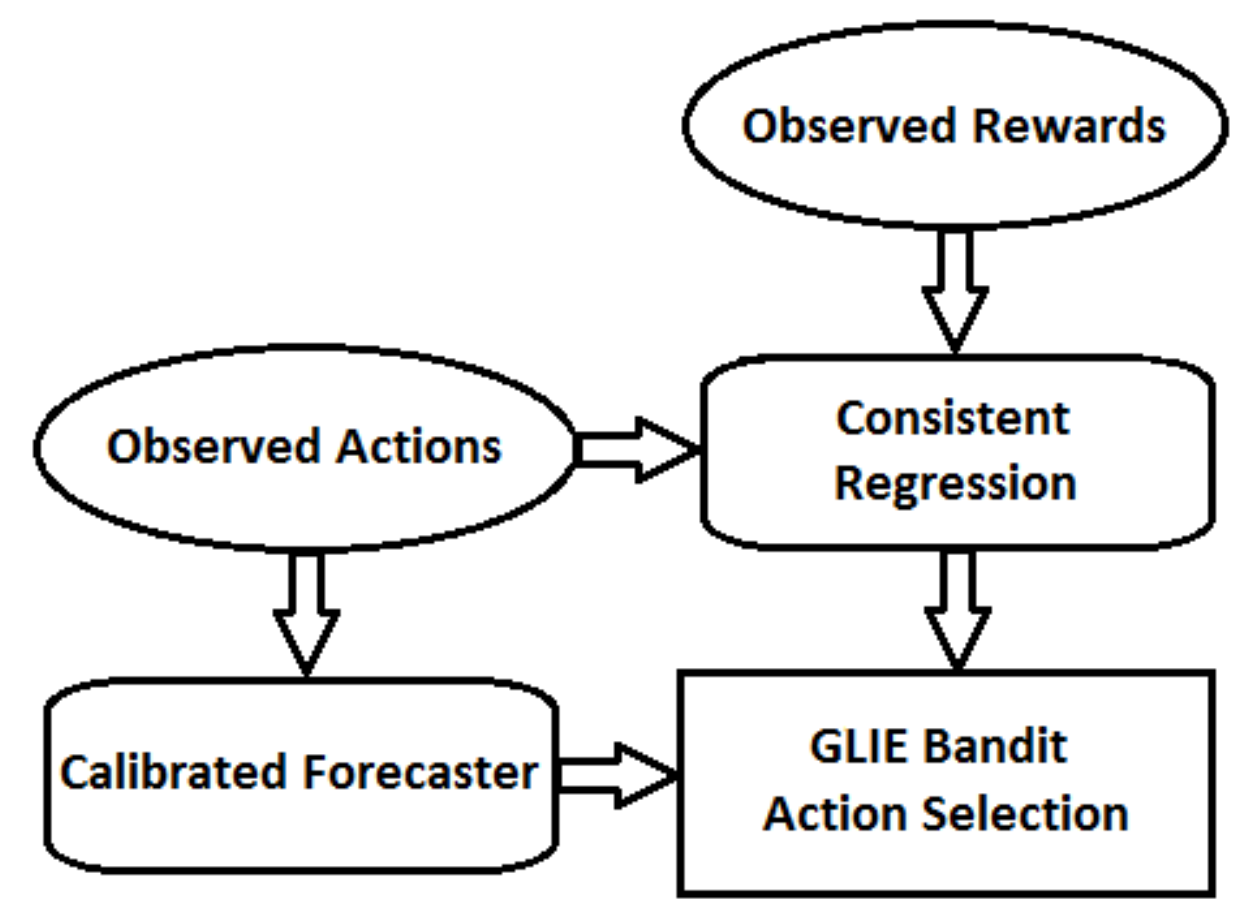}
\label{chart}}
\caption{Visualization of the proposed selection strategy.}
\label{fig:SysTrans}
\end{figure}

Note that in this approach, for larger period indices (large $r$), the fraction of time dedicated 
to exploration is smaller, as depicted in Figure \ref{EE}. Therefore, the strategy belongs to the 
class of algorithms that follow the \textit{''greedy in the limit with infinite exploration''} 
(GLIE) principal \cite{singh00}. Intuitively, this method is based on the fact that in a (near-) 
stationary environment, the estimation of reward processes of arms becomes more and more trust-worthy 
as time evolves, and therefore less exploration is required. Also note that the selection strategy 
and forecasting perform two different task; while the former refers to the estimation of reward 
processes, the latter predicts the joint action profile of opponents. The entire action selection 
procedure is visualized in Figure \ref{chart}.
\begin{algorithm}
\caption{Bandit Selection Strategy ($\chi$)}
\label{alg:Selection}
\small
\begin{algorithmic}[1]
\STATE Define an increasing sequence of integers, $T'_{r}=2^r$ for $r=1,2,...$. Each member 
$T'_{r}$ of this sequence denotes the length of period $r$, i.e. the number of trials included 
in it.
\STATE Define a decreasing sequence of numbers, $Z_{r}=\frac{r}{2^{r}}$ for $r=1,2,...$. 
\STATE Set the period $r=1$ and select the exploration parameter $\gamma \ll 1$.
\REPEAT
\STATE Select $\left \lceil T'_{r}\cdot Z_{r} \right \rceil$ exploration 
       trials belonging to $[1+\sum_{r}T'_{r-1},\sum_{r}T'_{r}]$ uniformly 
       at random.
\FOR{$t=s+\sum_{r}T'_{r-1}$, $1\leq s<T'_{r}$,}
  \IF{$t$ is an exploring trial,}
     \STATE 
        with probability $1-\gamma$, select an arm equally at random;\\
        with probability $\gamma$,\\
                1. receive the output of the forecaster (Algorithm 
                      \ref{alg:calibrated_forecaster}),\\
                2. using this information, select the arm with the highest 
                      estimated expected reward.            
  \ELSE
     \STATE Receive the input from the forecaster.
     \STATE Using this information, select the arm with the highest 
            estimated expected reward.
  \ENDIF
  \STATE Play the selected arm and observe the reward.
  \STATE Observe the actions of other players and inform the forecaster (forecaster's input).
  \STATE Re-estimate the mean reward function of the played arm.
\ENDFOR
  \STATE $r=r+1$.
\UNTIL{convergence ($r$ is sufficiently large)}
\end{algorithmic}
\end{algorithm}
%
\subsection{Strong-Consistency and Convergence}\label{subsec:Consist}

The following results ensure the consistency and declare the convergence characteristics 
of the proposed selection strategy.
\begin{lemma}
\label{lm:Assumption}
$T'_{r}=2^{r}$ and $Z_{r}=\frac{r}{2^{r}}$ satisfy Assumption \ref{as:bandit_game}.
\end{lemma}
\begin{IEEEproof}
The lemma can be easily verified by direct calculation using theorems concerning limits 
of infinite sequences.
\end{IEEEproof}
\begin{lemma}
\label{lm:Convergence}
Consider a selection strategy $\kappa$ so that each player $k$ plays with actions 
based on $\delta_{d_{t}}$ ($m_{k,t}=\kappa(\delta_{d_{t}})$), where $\delta_{d_{t}}$ 
is the Dirac probability distribution on the true joint action profile of its opponents 
at time $t$. Let $\kappa'$ be another strategy that is identical to $\kappa$, except 
that $P_{t}$ is used in the place of $\delta_{d_{t}}$ ($m_{k,t}=\kappa(P_{t})$), where 
$P_{t}$ is a probability distribution over all possible joint action profiles of opponents, 
produced by a calibrated forecaster. Then, $\lim_{T \to \infty} S_{\kappa,T} =1$ 
implies $\lim_{T \to \infty} S_{\kappa',T} =1$, where $S_{\kappa,T}$ and $S_{\kappa',T}$ 
are defined by (\ref{eq:consistency_metric}).
\end{lemma}
\begin{IEEEproof}
See Appendix \ref{subsec:lm_convergence}.
\end{IEEEproof}
Lemma \ref{lm:Convergence} simply states that if a strategy is strongly consistent 
given true joint action profiles, then its consistency is preserved by using the 
calibrated forecast of the joint action profiles.
\begin{lemma}
\label{lm:numberPlay}
Asymptotically, the selection strategy $\chi$ samples each action $m \in \left \{1,...,M \right \}$ 
and also each joint action profile $\textbf{m}=(m_{1},...,m_{K})\in \bigotimes_{k=1}^{K}\left \{1,...,M \right \}$ 
infinitely often.
\end{lemma}
\begin{IEEEproof}
See Appendix \ref{subsec:lm_numberPlay}.
\end{IEEEproof}
\begin{theorem}
\label{th:strong_consistent}
Under Assumptions \ref{as:expected_reward}, \ref{as:consistent_reg} and \ref{as:bandit_game}, 
the proposed selection strategy, $\chi$ (Algorithm \ref{alg:Selection}), is strongly-consistent.
\end{theorem}
\begin{IEEEproof}
See Appendix \ref{subsec:th_consistent}.
\end{IEEEproof}
\begin{theorem}
\label{th:Convergence}
Consider a K-player MAB game where each player is provided with $M$ actions. Let $\mathfrak{C}$ 
denote the set of correlated equilbria, $\textbf{m}=(m_{1},..,m_{K})\in \bigotimes_{k=1}^{K}\left \{ 1,..,M \right \}$, 
and define the empirical joint frequencies of play as (\ref{eq:empiricalFreq}). If all players 
play according to selection strategy $\chi$, then the distance $\inf_{\pi \in \mathfrak{C}}\sum_{\textbf{m}}\left | \hat{\pi}_{T}(\textbf{m})-\pi(\textbf{m})\right |$ between the empirical joint distribution of 
plays and the set of correlated equilibria converges to $0$ almost surely as $T \to \infty$.  
\end{theorem}
\begin{IEEEproof}
See Appendix \ref{subsec:th_convergence}.
\end{IEEEproof}
\begin{remark}
\label{re:jstifedAssumption}
In Section \ref{subsec:BanditModel}, we mentioned that every player is interested in optimizing its performance 
in the sense of regret minimization, and no player intends to ruin the performance of others. Therefore players 
are rational and not malicious. By Remark \ref{re:equivalency} and Theorem \ref{th:strong_consistent}, strategy 
$\chi$ yields vanishing regret; Thus the assumption that all players use this strategy is justified.
\end{remark}
\subsection{Some Notes on Convergence Rate}\label{subsec:Convergence}
\begin{figure}[ht]
\centering
\subfigure[Calibrated forecaster]{
\includegraphics[width=0.35\textwidth]{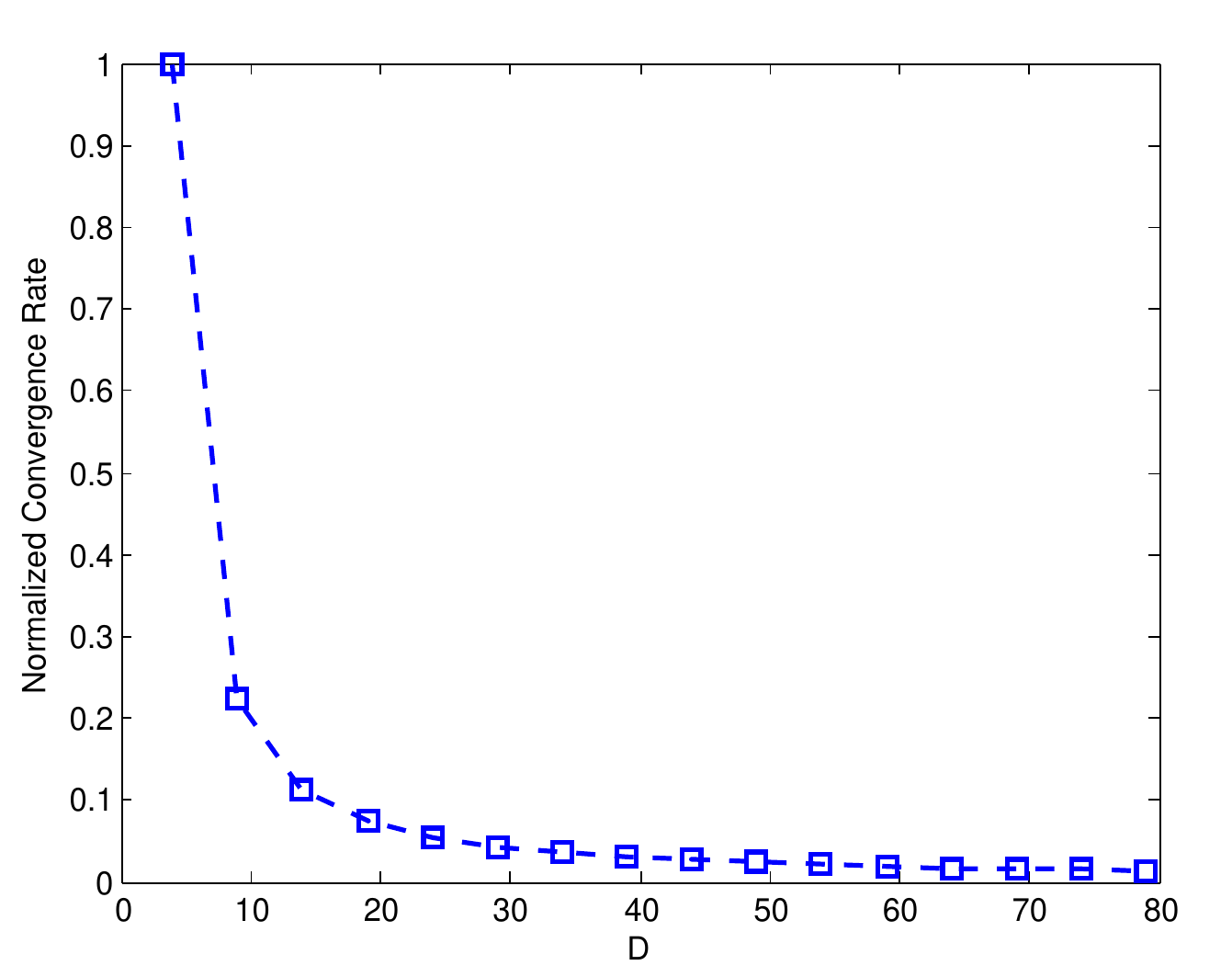}
\label{Fig:conFore}}
\quad
\subfigure[Regression]{
\includegraphics[width=0.35\textwidth]{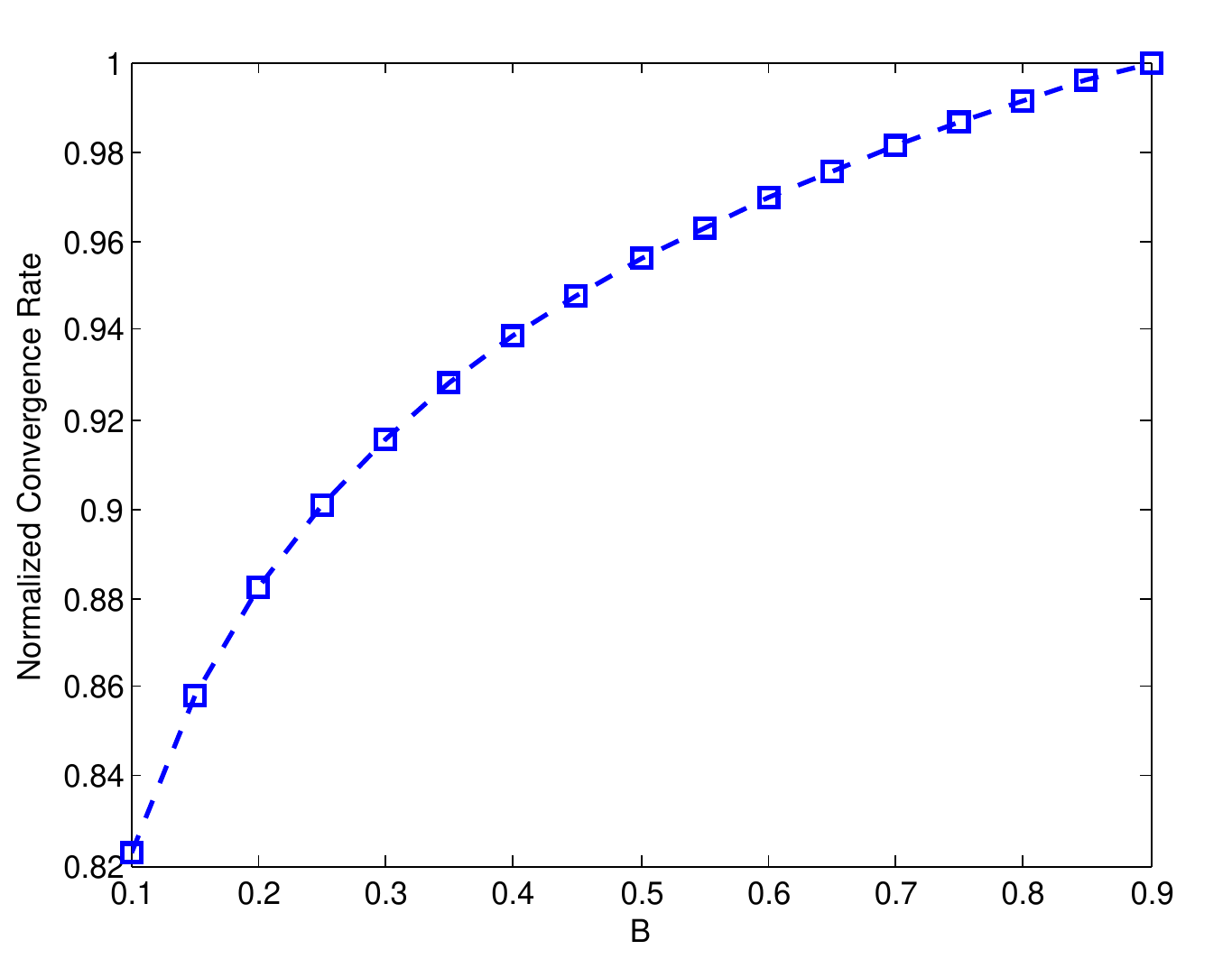}
\label{Fig:conReg}}
\caption{Scaling convergence rate with variables and parameters ($M$, $K$ and $\gamma$), $D=M^{(K-1)}>1$ and $B=\frac{1-\gamma}{M^{K}}<1$.}
\label{fig:Converg}
\end{figure}

As it is clear from Algorithm \ref{alg:Selection} (see also Figure \ref{chart}), for final convergence, the forecasting and 
regression procedures must converge to true joint action profile and true reward functions, respectively. In what follows, 
we discuss the impact of some variables, including number of actions ($M$) and users ($K$), as well as exploration parameter 
($\gamma$), on the convergence rate of these procedures.
\begin{theorem}[\cite{Mannor10}]
\label{th:RateForecat}
For the calibrated forecaster given in Algorithm \ref{alg:calibrated_forecaster} we have
\begin{equation}
\limsup_{T\to \infty } \frac{T^{\frac{1}{D+1}}}{\sqrt{\ln (T)}} \sup_{B \in \mathfrak{B}}\left \| \frac{1}{T}\sum_{t=1}^{T} \mathbb{I}_{P_{t} \in B}(P_{t}-\delta _{d_{t}})\right \|_{1}\leq \Gamma_{D},
\end{equation}
where $\mathfrak{B}$ is the Borel sigma-algebra of $\mathfrak{P}$ and the constant $\Gamma_{D}$ depends only on $D$.
\end{theorem}
From the algorithm we know that $D=M^{(K-1)}>1$. Figure \ref{Fig:conFore} shows how the convergence rate scales 
with $D$ for $\Gamma_{D}=D$. As expected, convergence speed decreases for larger number of users $K$ and/or actions 
$M$, thereby larger $D$. Note that the effect of increasing $K$ on $D$ is more than that of $M$. 

Now, consider the regression process, which is assumed to be non-parametric. Then the following holds.
\begin{theorem}[\cite{Stone82}]
\label{th:RateReg}
Consider a $p$-times differentiable unknown regression function $f$ and a $d$-dimensional measurement variable. Let $\hat{f}$ 
denote an estimator of $f$ based on a training sample of size $n$, and let $\left \|\hat{f}_{n}-f \right \|_{\infty}$ be the 
$L_{\infty}$ norm $\hat{f}_{n}-f$. Under appropriate regularity condition, the optimal rate of convergence for $\left \|\hat{f}_{n}-f \right \|_{\infty}$ 
to zero is $\left (\frac{\log(n)}{n} \right)^{\eta}$ where $\eta=\frac{p}{2p+d}$.
\end{theorem}
Based on Theorem \ref{th:RateReg}, Algorithm \ref{alg:Selection} changes the convergence rate through changing the sampling 
rate. Assume that samples are gathered only at exploration trials. Let $R$ be the number of periods (game horizon). By 
the algorithm, each joint action profile is expected to be played $\frac{1-\gamma}{M^{K}}\sum_{r=1}^{R}r=\frac{1-\gamma}{M^{K}}\cdot\frac{R(R+1)}{2}$ 
times during $R$ periods (see also the proof of Lemma \ref{lm:numberPlay}). Moreover, suppose that some fixed number of 
samples are required to estimate the reward of each joint action profile with some precision. Therefore, it is clear that 
increasing $M$ and/or $K$, as well as increasing $\gamma$, degrades the sampling rate and thereby the convergence speed, 
since larger game horizon is required for sufficient sampling. Let $B=\frac{1-\gamma}{M^{K}}<1$. Figure \ref{Fig:conReg} 
shows how changes in $B$ impact the convergence speed of regression process.
%
%
\section{Numerical Results}\label{sec:Numerical}
This section consists of two parts. First, we consider a simple model, and clarify how the algorithm works. 
Next we consider some larger network and the performance of the proposed approach is compared with some other 
approaches.  
\subsection{Part One}\label{sec:NumericalOne}
\subsubsection{Network model}\label{subsec:SimScenario}
We consider an underlay D2D network consisting of two D2D users ($K=2$). We assume that there exist two 
primary channels ($M=2$), whose availability follows Bernoulli distribution with parameter $\frac{1}{2}$. 
We implement the following selection strategies.
\begin{itemize}
\item Statistical centralized strategy (SC): Given global \textit{statistical} channel knowledge and by exhaustive 
      search, a central controller assigns each D2D user some transmission channel so that the assignment corresponds 
      to the most efficient pure strategy equilibrium point in the sense of maximum aggregate average throughput.
\item Calibrated bandit strategy (CB): Provided with no prior information, D2D users simultaneously utilize 
      the selection strategy $\chi$, described in Algorithm \ref{alg:Selection}. 
\end{itemize}

Since $M=2$, for each D2D user $k \in \left \{1,2 \right \}$ we have $p_{k,1}+p_{k,2}=1$, where $p_{k,i}$ is the likelihood 
of D2D user $k$ to take action $i \in \left \{1,2 \right\}$ by following the mixed strategy $(p_{k,1},p_{k,2})$. This implies 
that there exists only one degree of freedom in the $\epsilon$-grid of forecasters, i.e. for each player $k$ the probability 
distribution over all joint action profiles of opponents reduces here to the mixed strategy of the other player. We assume 
$N_{\epsilon}=40$.\footnote{In general, smaller $N_{\epsilon}$ can be used at early periods to reduce the computational burden. 
In this example, however, we fix $N_{\epsilon}$ for all periods in order to highlight the evolution of outputs over time.}~Therefore, 
the $\epsilon$-grid defines 40 possible mixed strategies (quantized vectors).\footnote{Vectors are indexed as $i=1,...,40$. For 
$i=1$, $(p_{1},p_{2})=(0,1)$, while for $i=40$, $(p_{1},p_{2})=(1,0)$. That is, $p_{1}$ increases with the index of quantized 
vector, while $p_{2}$ decreases.}~The primal output of the forecaster of a player is a vector of weights including 40 elements, 
where each element denotes the likelihood of one of the \textit{quantized} mixed strategies to be played by the other player. 
The final output of the forecaster is then a mixed strategy extracted from the set of quantized mixed strategies according to 
this distribution, as described in Algorithm \ref{alg:calibrated_forecaster}. 
\subsubsection{Orthogonal Multiple Access}\label{subsec:Ortho}
\begin{figure}[t]
\centering
\includegraphics[width=0.4\textwidth]{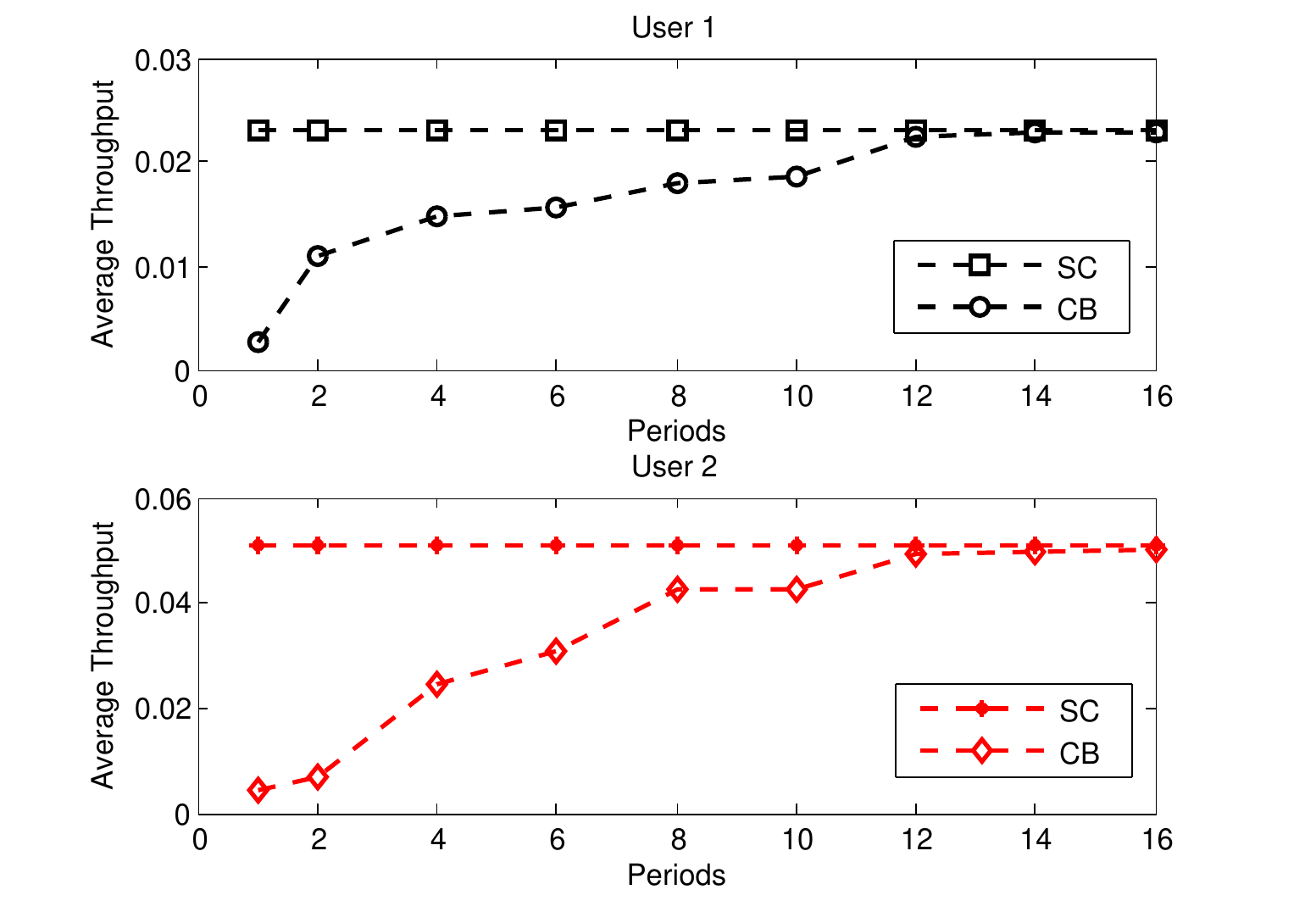}
\caption{Average throughput of our approach versus that of centralized strategy (orthogonal access).}
\label{TgOrth}
\end{figure}

\begin{figure}[t]
\centering
\includegraphics[width=0.3\textwidth]{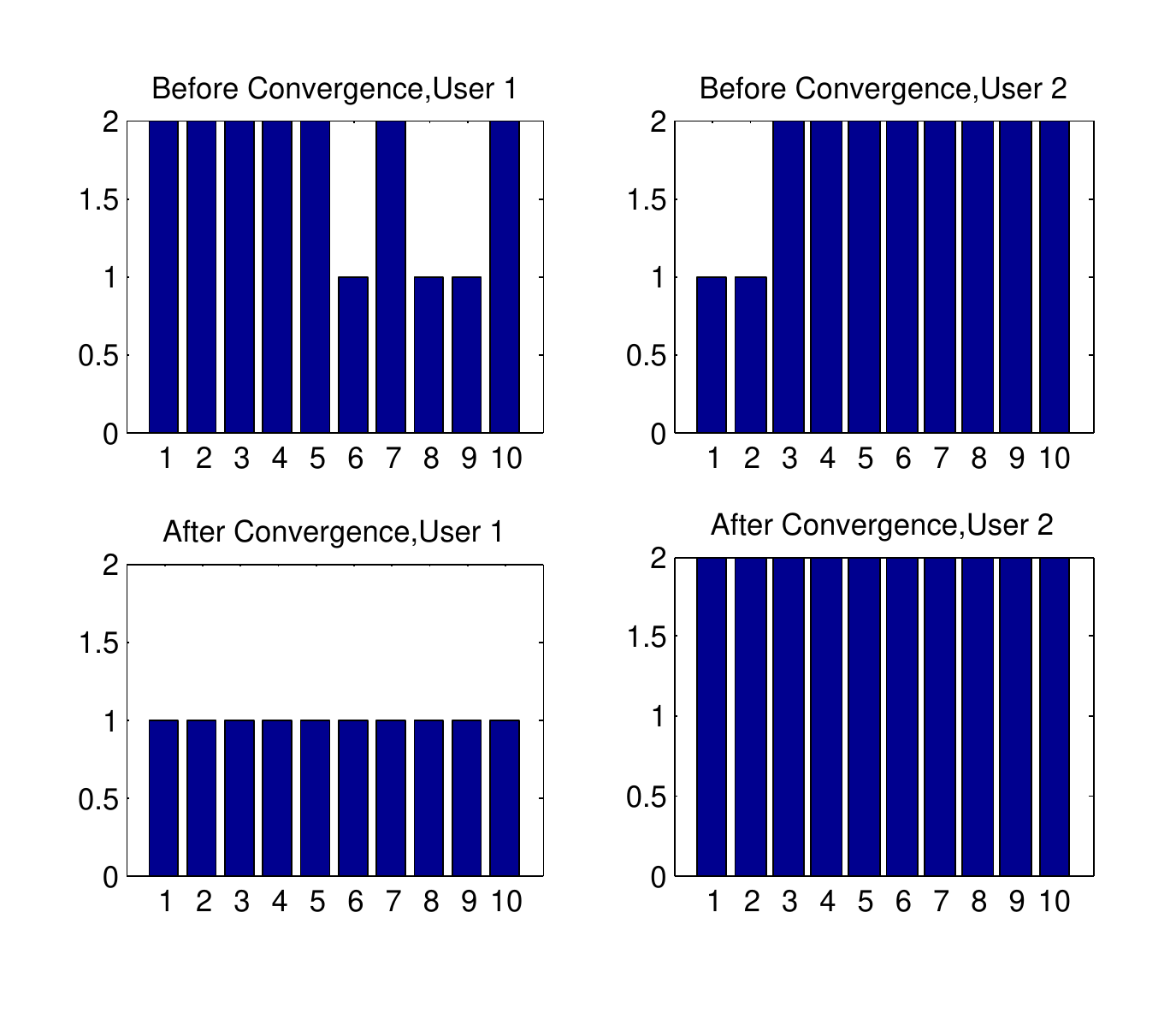}
\caption{Selected actions before and after convergence (orthogonal access).}
\label{ActBeforeAfter}
\end{figure}

\begin{figure}[ht]
\centering
\subfigure[Before convergence.]{%
\includegraphics[width=0.4\textwidth]{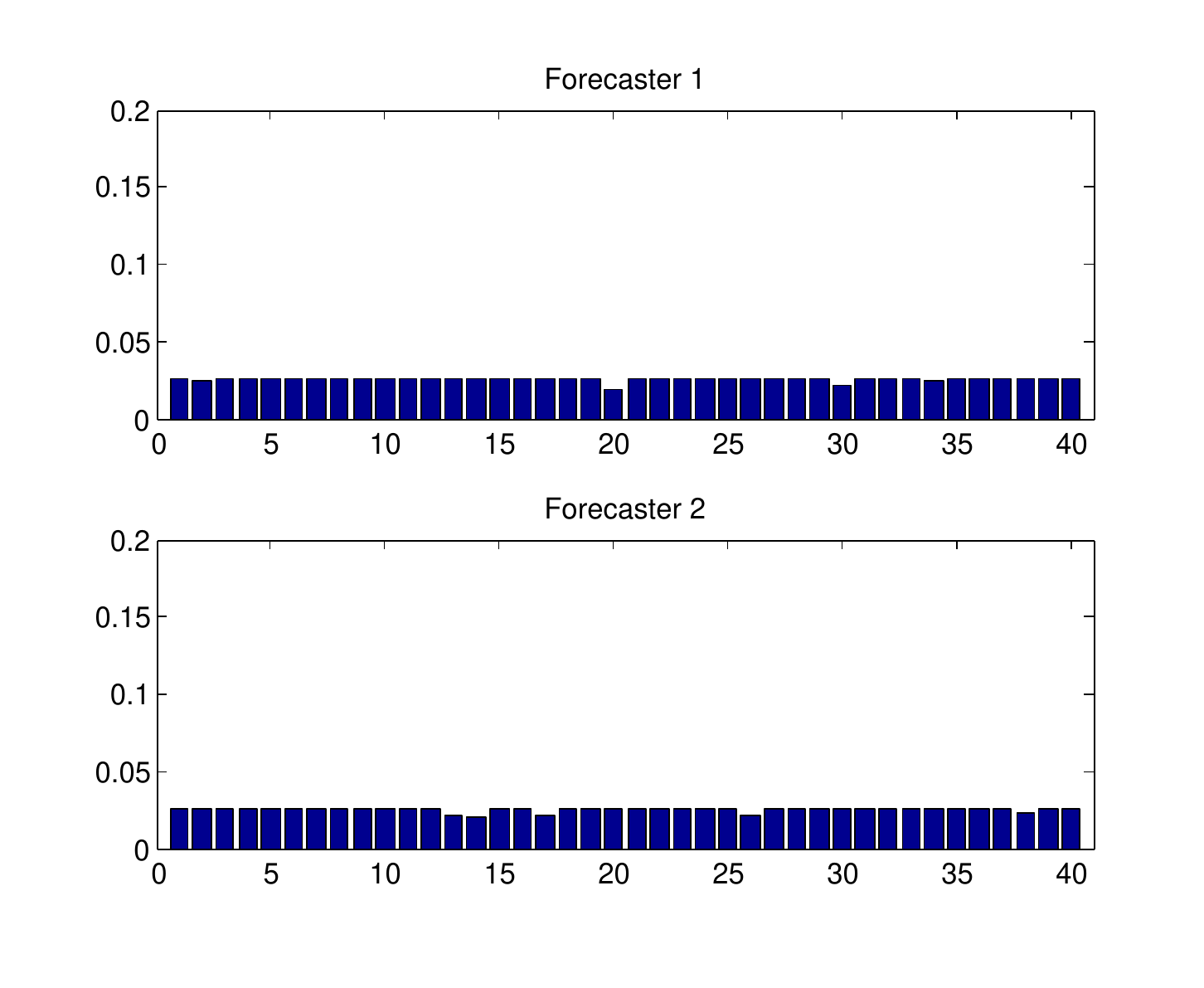}
\label{BeForOrth}}
\quad
\subfigure[After convergence.]{%
\includegraphics[width=0.4\textwidth]{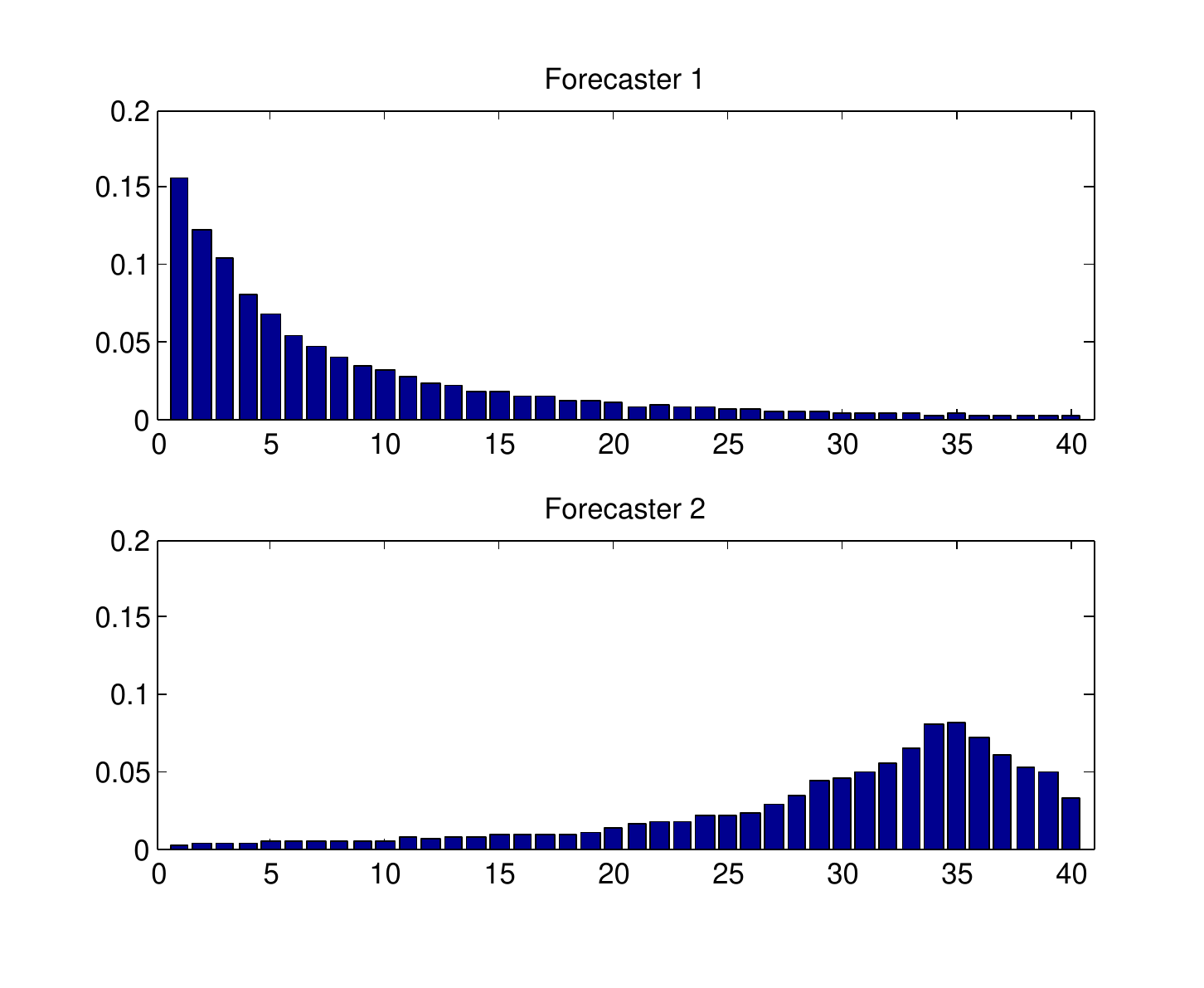}
\label{AftForOrth}}
\caption{Forecasters' outputs (orthogonal access).}
\label{fig:ForecastOrtho}
\end{figure}

Assume that D2D users follow the orthogonal transmission scheme, described in Section \ref{sec:System}. 
Based on average channel gains, the joint rewards of players under possible joint action profiles are 
summarized in Table \ref{Tb:Orthogonal}. From this table, the channel selection game has a pure-strategy 
Nash equilibrium\footnote{Note that Nash equilibrium is a special case of correlated equilibrium.}~that 
yields the maximum aggregate reward for the two D2D users, and is achieved when D2D users 1 and 2 transmit 
in the first and second channels respectively (i.e. joint action (1,2)).\\ 
\begin{table}[h]
\caption{Joint reward table (orthogonal access)}
\label{Tb:Orthogonal}
\begin{center}
  \begin{tabular}{ |c | c | c| }
  \hline
  channel    & $1$           &  $2$            \\ \hline 
     $1$     & 0.012,0.000   &  0.023,0.054    \\ \hline
     $2$     & 0.016,0.000   &  0.008,0.027    \\ \hline
  \end{tabular}
 \end{center}
\end{table}  
The average throughput achieved by each player is depicted in Figure \ref{TgOrth}. It can be seen that 
for sufficiently large game horizon (number of periods), the average throughput of our strategy converges 
to that of equilibrium.
Actions of players are shown in Figure \ref{ActBeforeAfter}, at both early and final 
stages of the game, i.e. before and after the convergence, for 10 consecutive trials. 
By comparing this figure with the data given in Table \ref{Tb:Orthogonal}, it follows 
that the game converges to equilibrium, which is the joint action $(1,2)$. Moreover, 
the primal outputs of forecasters ($\psi$, see Algorithm \ref{alg:calibrated_forecaster}) 
are shown in Figure \ref{BeForOrth}, for some trial before convergence. In this figure, 
outputs are almost uniformly distributed, meaning that all quantized mixed strategies 
are almost equally likely to occur. This result is in agreement with Figure \ref{ActBeforeAfter}, 
where selected channels before convergence do not follow any specific pattern. On the 
other hand, outputs of forecasters at some trial after convergence are depicted in 
Figure \ref{AftForOrth}. In this figure, Forecaster 1 assigns higher weights to quantized 
mixed strategies with $p_{2}>p_{1}$, while Forecaster 2 emphasizes the strategies with 
$p_{1}>p_{2}$. This means that first and second players are excepted to select channels 1 
and 2, respectively, by their opponents. These predictions are again approved by Figure \ref{ActBeforeAfter}, 
where first and second D2D users finally settle at first and second channels, respectively.  

\subsubsection{Non-orthogonal Multiple Access}\label{subsec:Northo}
As described in Section \ref{sec:System}, in case of non-orthogonal multiple access, conflicting D2D 
users transmit simultaneously, which might result in interference. In this scenario, players decide 
whether to solve the conflict by diverting to different channels, or to transmit in a common channel. 
In order to clarify this, we perform two experiments. For the first and second experiments, joint 
rewards are given by Table \ref{Tb:NorthoOne} and Table \ref{Tb:NorthoTwo}, respectively.
%
\begin{table}[ht]
\centering
\subtable[Case 1]{
\begin{tabular}{ |c | c | c| }
\hline
channel    & $1$           &  $2$        \\ \hline 
$1$     & 0.024,0.040   &  0.024,0.021   \\ \hline
$2$     & 0.075,0.042   &  0.063,0.021   \\ \hline
\end{tabular}
\label{Tb:NorthoOne}
}
\quad
\subtable[Case 2]{
\begin{tabular}{ |c | c | c| }
\hline
channel    & $1$           &  $2$         \\ \hline 
$1$     & 0.024,0.001   &  0.024,0.021    \\ \hline
$2$     & 0.075,0.000   &  0.063,0.021    \\ \hline
\end{tabular}
\label{Tb:NorthoTwo}
}
\caption{Joint reward table (non-orthogonal access)}
\end{table}
From these tables, the most efficient pure-strategy equilibrium points for the first and second 
games are joint actions $(2,1)$ and $(2,2)$ respectively. This means that in the first case, it 
is beneficial for players transmit in different channels, while in the second case, D2D users 
achieve higher gains if both transmit through the second channel.

\begin{figure}[ht]
\centering
\subfigure[Case 1]{
\includegraphics[width=0.4\textwidth]{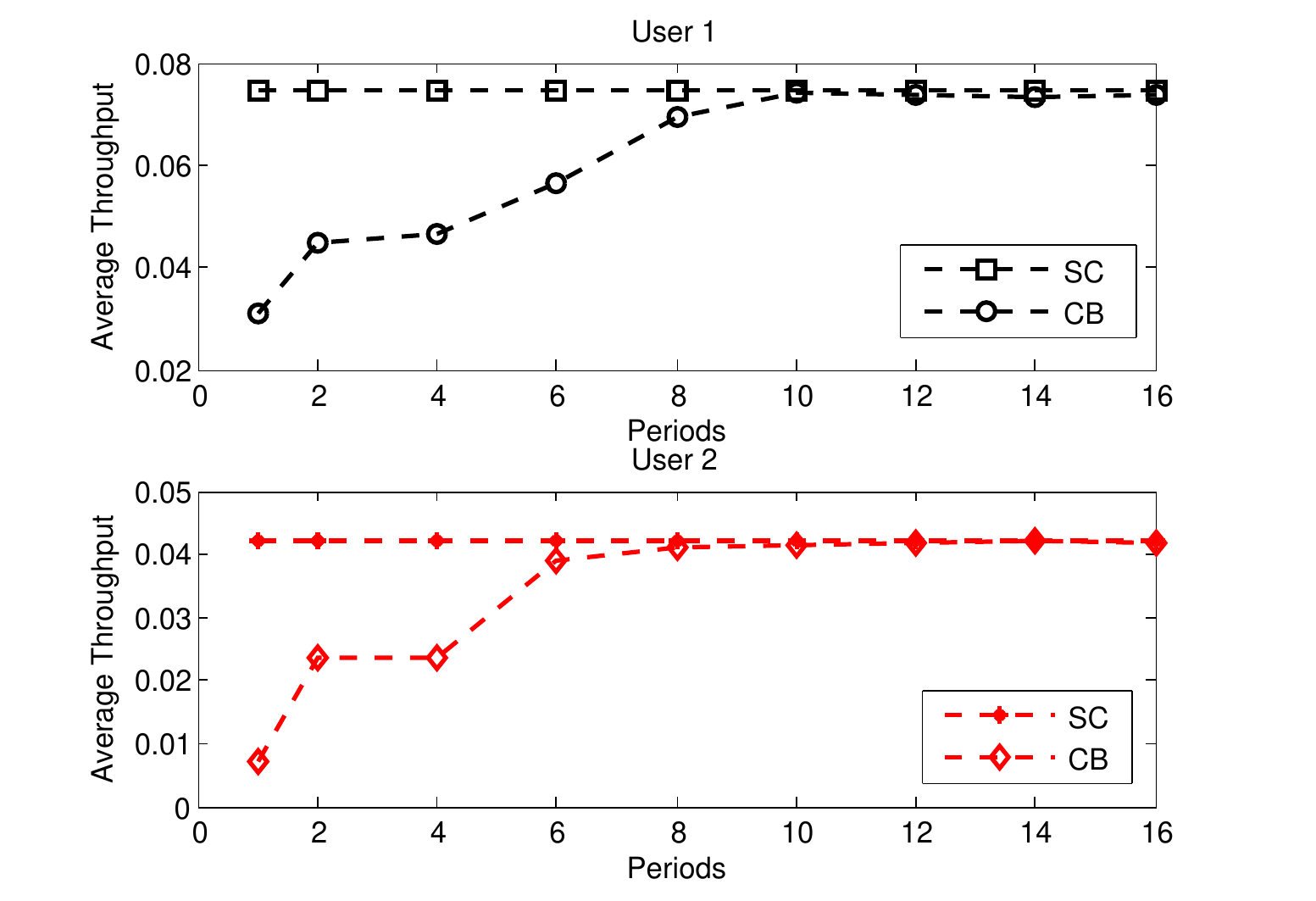}
\label{TgNonOne}}
\quad
\subfigure[Case 2]{
\includegraphics[width=0.4\textwidth]{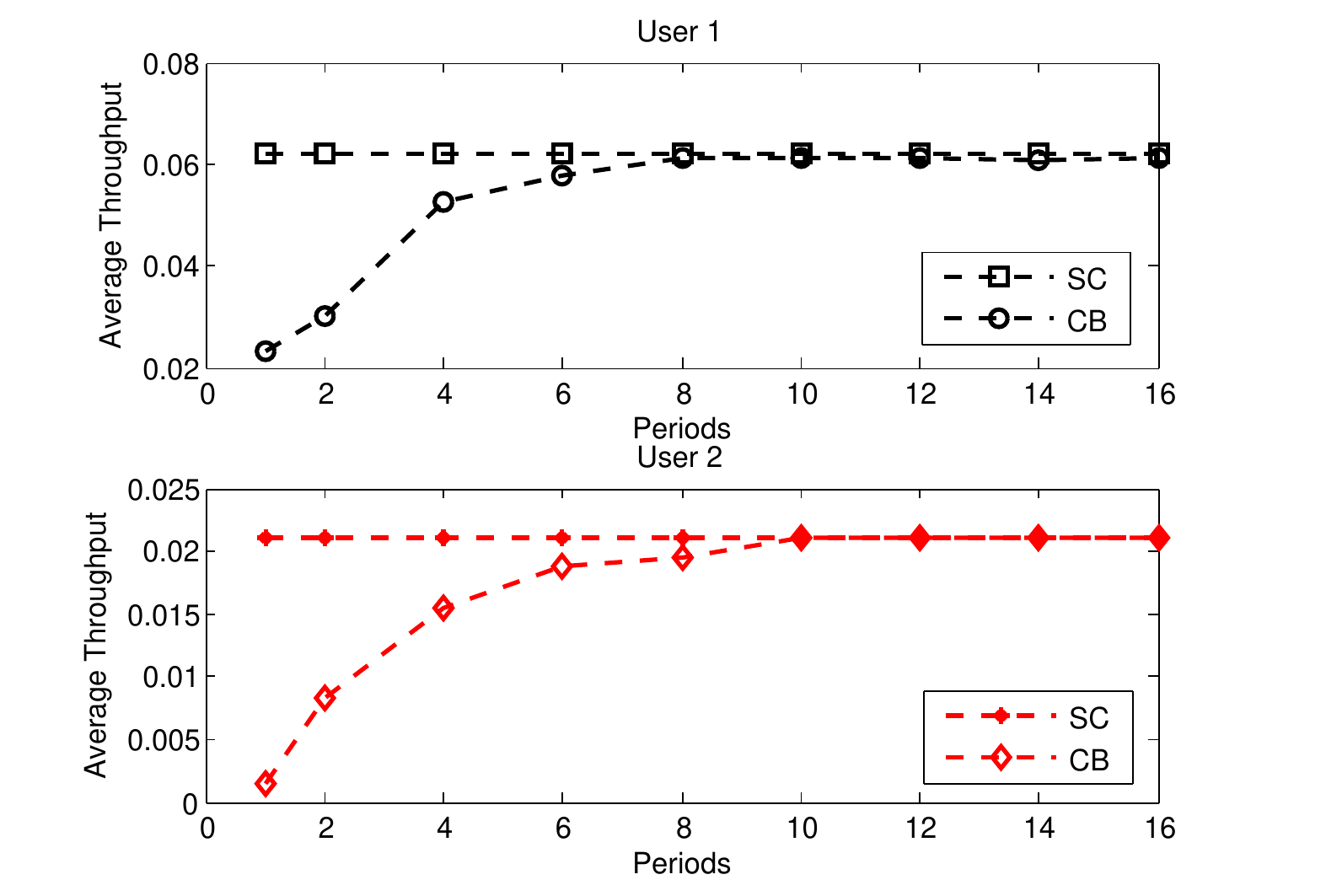}
\label{TgNonTwo}}
\caption{Average throughput of our approach versus that of equilibrium (non-orthogonal 
access).}
\label{fig:AverageTg}
\end{figure}

\begin{figure}[ht]
\centering
\subfigure[Case 1]{
\includegraphics[width=0.30\textwidth]{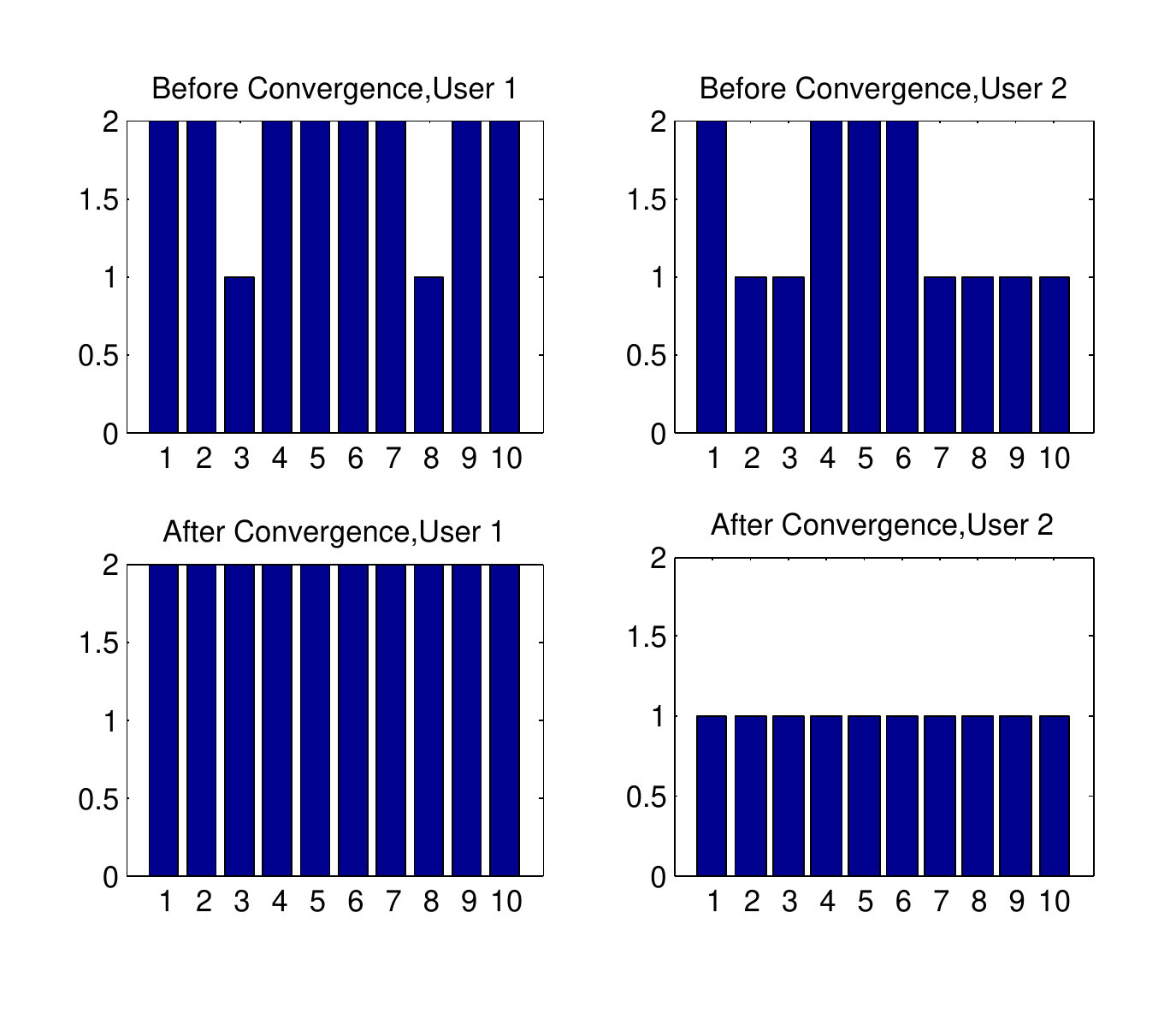}
\label{ActNonOneBeAft}}
\quad
\subfigure[Case 2]{
\includegraphics[width=0.30\textwidth]{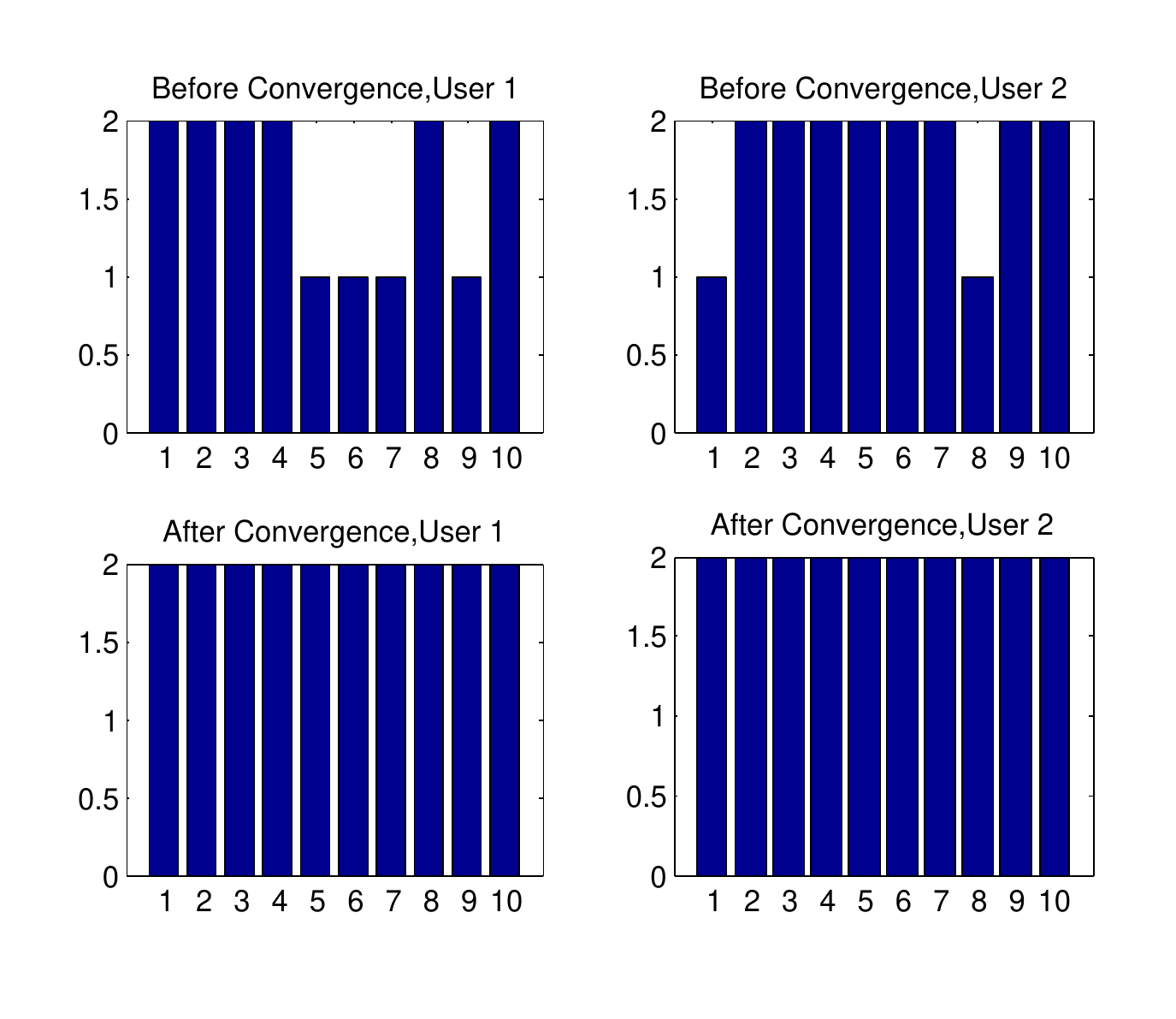}
\label{ActNonTwoBeAft}}
\caption{Selected actions before and after convergence (non-orthogonal access).}
\label{fig:ActNortho}
\end{figure}

\begin{figure}[ht]
\centering
\subfigure[Case 1]{
\includegraphics[width=0.4\textwidth]{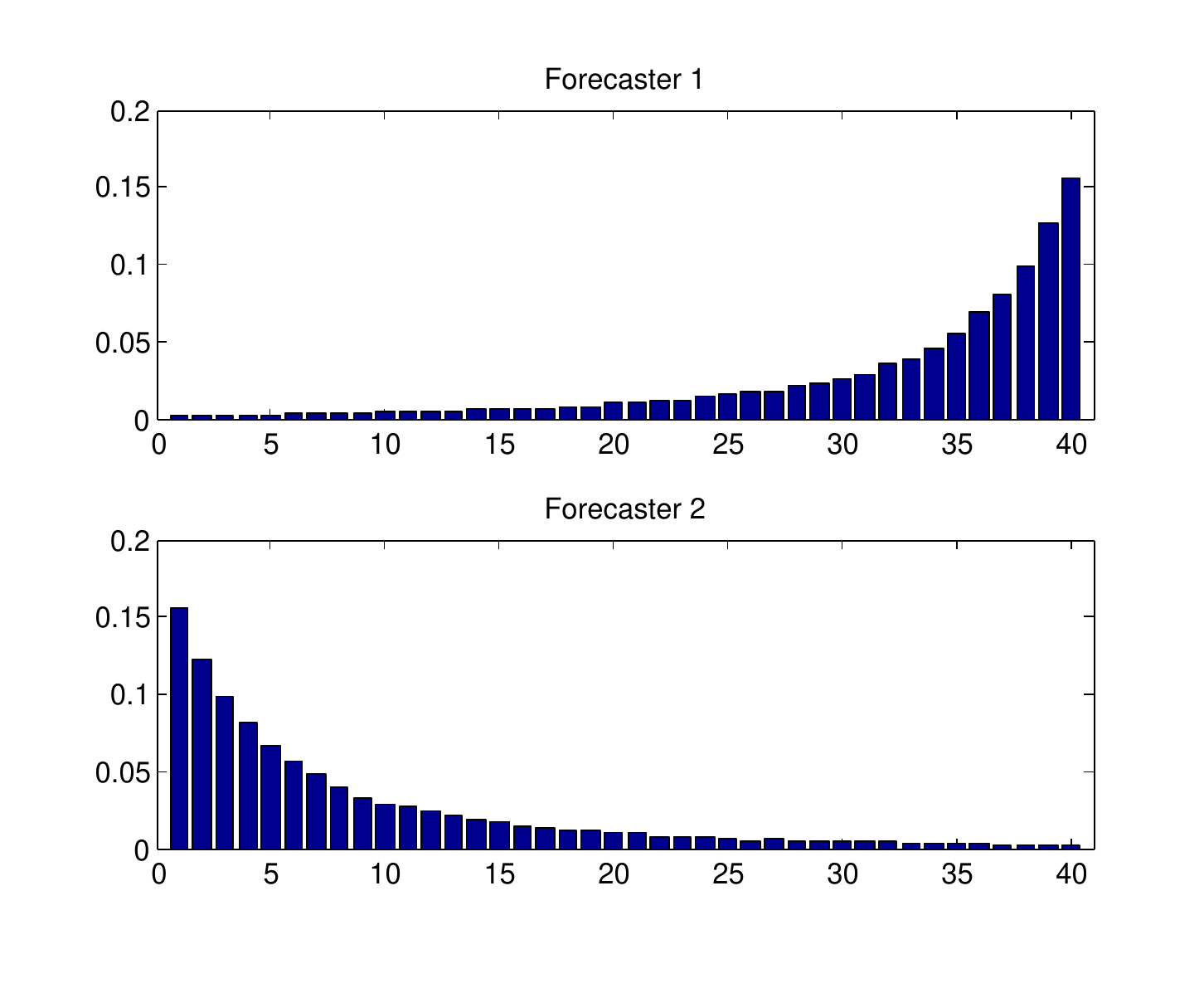}
\label{AftForNonOne}}
\quad
\subfigure[Case 2]{
\includegraphics[width=0.4\textwidth]{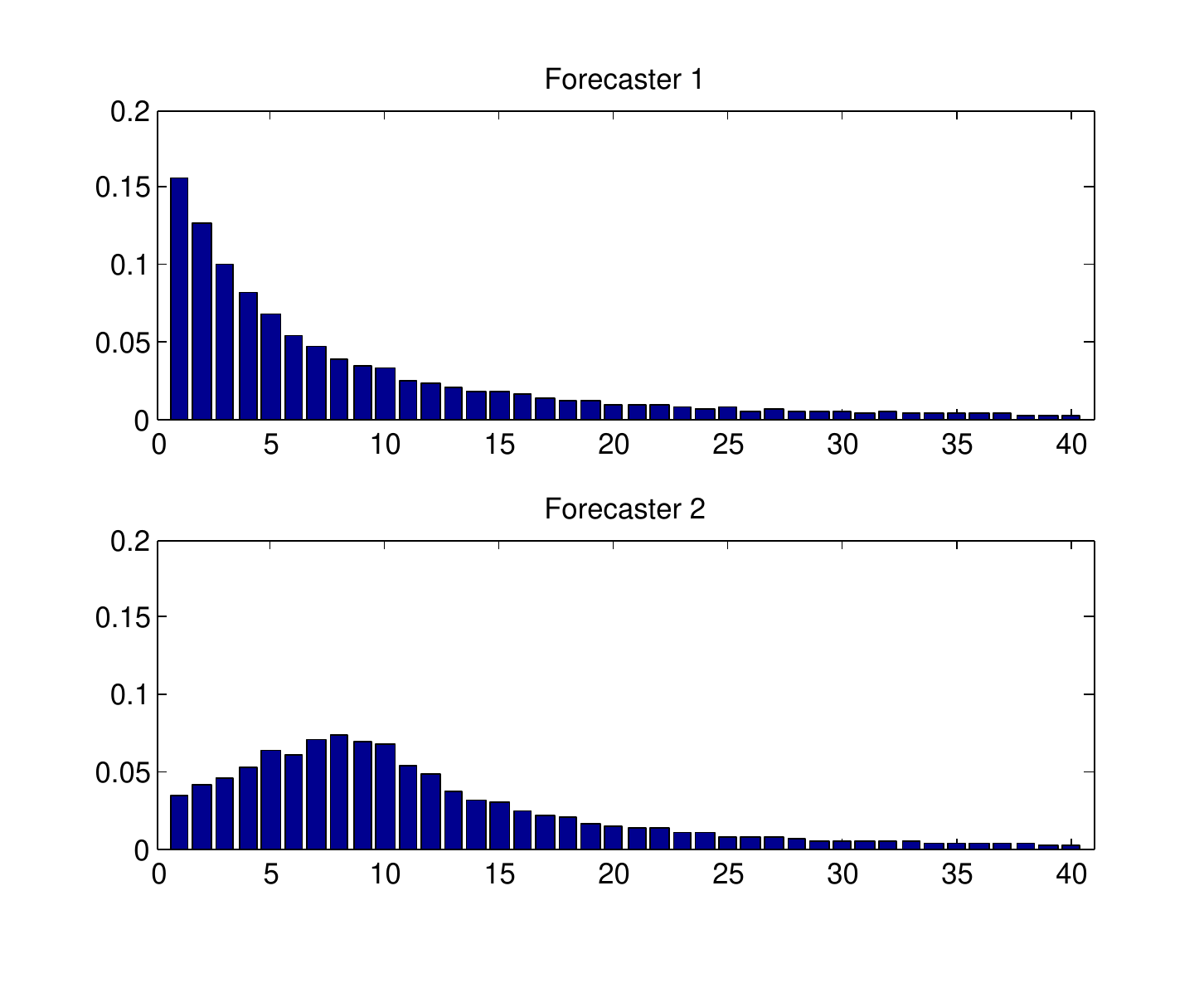}
\label{AftForNonTwo}}
\caption{Forecasters' outputs \textit{after} convergence (non-orthogonal access).}
\label{fig:ForecastNortho}
\end{figure}

The achieved throughput by players are shown in Figures \ref{TgNonOne} and \ref{TgNonTwo}, 
respectively for the two experiments. Moreover, Figures \ref{ActNonOneBeAft} and \ref{ActNonTwoBeAft} 
show the actions of players for first and second experiments. Figures \ref{AftForNonOne} 
and \ref{AftForNonTwo}, at a single trial after convergence (the outputs of forecasters 
before convergence are similar to Figure \ref{BeForOrth}). Descriptions are similar to 
the orthogonal case, and are omitted for space considerations.

%
\subsection{Part Two}\label{sec:NumericalTwo}
\begin{figure}[t]
\centering
\includegraphics[width=0.50\textwidth]{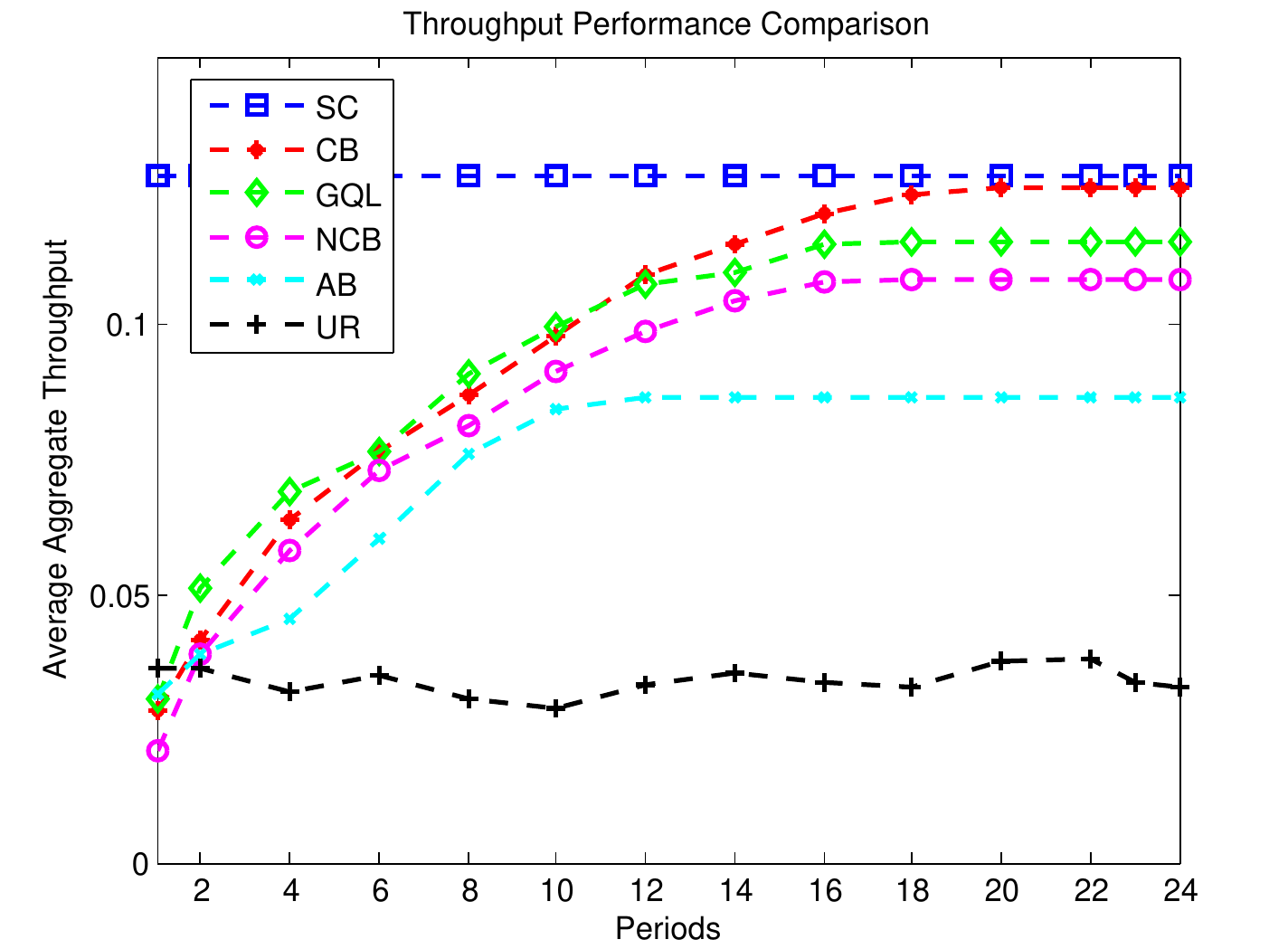}
\caption{Throughput performance comparison.}
\label{Fig:Comp}
\end{figure}

Consider a D2D network with 4 users and 4 primary channels. The performance metric is the aggregate average throughput of users, 
and the following approaches are compared.
\begin{itemize}
\item Statistical centralized strategy (SC) : This approach is described in Section \ref{subsec:SimScenario}. 
\item Calibrated bandit strategy (CB): This is our proposed selection strategy (Algorithm \ref{alg:Selection}).
\item No collision bandit strategy (NCB): Following \cite{KalathilD12}, stochastic multi-armed bandit game is played, where in case 
      of collision, \textit{no} reward is assigned to colliding users. 
\item $\epsilon$-greedy Q-learning strategy (GQL): Let $0<\epsilon<1$. Each player assigns some Q-value to each action-state pair. 
      At each trial, every player selects the action that has yield the largest Q-value so far with probability $1-\epsilon$, and an 
      action uniformly at random with probability $\epsilon$. After playing, the Q-value of the selected action and observed state is 
      updated \cite{Bennis11}. Note that \textit{no} forecasting is performed, thus the best-response dynamics cannot be applied. 
\item Availability-based Strategy (AB) : As described in \cite{XuA12}, this model ignores the role of channel qualities in the reward
      achieved by players. More precisely, the learning approach includes only the availability of channels and the number of users 
      willing to transmit through each channel.  
\item Uniformly Random Strategy (UR): At each trial, an action is selected uniformly at random.
\end{itemize}
Results are depicted in Figure \ref{Fig:Comp}, and discussed briefly in the following.
\begin{itemize}
\item CB requires some time to converge to the average throughput achieved by SC. It yields no overhead, however, since unlike SC, 
      D2D users are not required to establish direct contact with the BS. Computational effort is also much less than that of SC.
\item The performance of GQL algorithm is inferior to the performance of CB. This is mainly due to the absence of forecasting and 
      best-response dynamics. Note that in addition to aggregate performance loss, simple $\epsilon$-greedy algorithms without 
      forecasting do not guarantee that the game converge to an equilibrium.         
\item The main reason that NCB performs poor is that the collision is not allowed. In such condition, even if it is better for D2D 
      users to collide (similar to the result of Section \ref{subsec:Northo}, case 2), the approach makes them to choose different 
      channels. 
\item Similar to GQL, AB does not exhibit good performance in comparison to CB. The reason is obvious. While CB takes channel quality 
      and availability into account, AB is only based on channel availability. Intuitively, the performance of AB is in direct 
      relation with channel qualities. More precsiely, for channels with similar and/or large gains, the harmful effect of ignoring 
      channel qualities is alleviated. The advantagemof this approach is its zero overhead.  
\item Uniform random strategy yields the worst performance. However, it is also the simplest approach, with respect to information flow 
      and computational effort. 
\end{itemize}
%
 
%
\section{Conclusion and Remarks}\label{sec:conclusion}
We studied a channel selection problem in an underlay distributed D2D communication system. 
In this model, spectrum vacancies of the cellular network are utilized by D2D users, thereby 
boosting the spectrum efficiency and improving local services, with no adverse effect on 
cellular users. Each D2D user aims at maximizing its performance given no prior information, 
and achieving an equilibrium is beneficial for all users. We showed that the channel selection 
problem boils down to a multi-player multi-armed bandit game with side information, and we 
proposed an approach to solve this game by combining no-regret bandit learning with calibrated 
forecasting. Analytically, we established that the proposed strategy is strongly-consistent; 
that is, for each D2D user, the average accumulated reward in the long run is equal to that 
based on the best fixed strategy in te sense of aggregate average reward. Moreover, we proved 
that the proposed approach converges to an equilibrium in some sense. Numerical analysis verified 
analytical results.

%
\section{Appendix}\label{sec:Appendix}
%
\subsection{Proof of Lemma \ref{lm:Convergence}}\label{subsec:lm_convergence}
We follow a root suggested in \cite{Chapman11}. 
Suppose that $\lim_{T \to \infty} S_{\kappa,T} =1$ holds. In order to prove 
$\lim_{T \to \infty} S_{\kappa',T} =1$, it is sufficient to show that after some 
finite time, the actions taken by the player based on $\mathbf{P}_{t}$ are equal 
to those based on $\delta_{d_{t}}$. By (\ref{eq:calibration}), we know that, with 
probability 1, there exists an $\nu>0$ so that after a time point $\theta<\infty$, 
\begin{equation}
\label{eq:Cons-Lm}
\left |\mathbf{P}_{t}-\delta_{d_{t}} \right|< \nu
\end{equation}
holds for all $t>\theta$ (see also Theorem \ref{th:CalibAlgorithm}). At the 
same time, according to our system model and by Assumption \ref{as:expected_reward}, 
the reward functions are bounded, and the action space and memory are finite. This 
implies that if (\ref{eq:Cons-Lm}) holds, then the actions of the player evolve as 
if it were aware of the true joint action profile of its opponents. Hence 
the lemma follows.
\subsection{Proof of Lemma \ref{lm:numberPlay}}\label{subsec:lm_numberPlay}
From Algorithm \ref{alg:Selection}, at each period $r$, $\left \lceil T'_{r}\cdot Z_{r}
\right \rceil$ trials are selected for exploration by each player. At each one of these trials, with 
probability $1-\gamma$, an arm $m$ is selected equally at random. Since these processes 
are independent, the probability that arm $m$ is pulled at some exploration trial yields 
$\frac{1-\gamma}{M}$. Now, let $\left \{W_{t}^{(m)}\right \}_{t=1}^{T}$ be a sequence 
of random variables, where $W_{t}^{(m)}=1$ if arm $m$ is played at time $t$, and $W_{t}^{(m)}=0$ 
otherwise, and the outcomes $W_{t}^{(m)}$ are independent over time. In the worst-case, 
arm $m$ never becomes the best-response, and hence its chance of being played is limited 
to exploration trials. As a result, $\textup{Pr}\left [W_{t}^{(m)}=1 \right ]=\frac{1-\gamma}{M}$, 
and the sum of probabilities for the event $W_{t}^{(m)}=1$ yields $ \sum_{t=1}^{T}\textup{Pr}\left [W_{t}=1 \right ]= \frac{1-\gamma}{M}\sum_{r=1}^{R}\left \lceil T'_{r}\cdot Z_{r}\right \rceil$. By using 
Assumption \ref{as:bandit_game} and Lemma \ref{lm:Assumption}, 
we conclude that $ \lim_{T \to \infty}\sum_{t=1}^{T}\textup{Pr}\left [W_{t}=1 \right ]\to\infty$. 
Thus, by the second Borel-Cantelli lemma (\cite{Chapman11}, \cite{feller68}), it follows 
that the probability of arm $m$ being pulled infinitely often equals 1. On the other hand, 
players select their actions independently. As such, the probability of playing each joint 
action profile is given by $\frac{1-\gamma}{M^{K}}$. By the same argumentation, each joint 
action profile is also played infinitely often. Hence, the Lemma is proved.

\subsection{Proof of Theorem \ref{th:strong_consistent}}
\label{subsec:th_consistent}
Since the proof is identical for all players, we prove the 
strong-consistency for some player $k$, and hence omit the player's subscript, 
$k$, for brevity. It should be mentioned that the proof is inspired by \cite{Yang02}, 
where the authors showed the consistency of an allocation rule for single-player 
contextual bandit games, which, similar to our algorithm, follows the GLIE rule. 

In the following, we consider a selection strategy $\chi'$, which is identical to 
$\chi$, except that at each time $t$ and before taking any action, the player is 
informed about the \textit{true} joint action profile of other $K-1$ players, that 
is $k^{-}_{t}$. We prove that $\chi'$ is strongly-consistent. Therefore, from 
Lemma \ref{lm:Convergence} it follows that $\chi$ is strongly-consistent, as well.  

In what follows, $m_{t}$ is used to denote the selected arm at time $t$, 
while $\hat{m}_{t}$ stands for the arm with the highest \textit{estimated} expected reward at 
time $t$. That is, at time $t$ we have $\hat{f}_{\hat{m}_{t}}(k^{-}_{t})\coloneqq \max_{m \in \left \{1,..,M \right\}} \hat{f}_{m}(k^{-}_{t})$. 
Moreover, $m^{*}_{t}$ denotes the arm with the highest \textit{true} expected reward at time 
$t$ so that $f_{m^{*}_{t}}(k^{-}_{t})\coloneqq \max_{m \in \left \{1,..,M \right\}} f_{m}(k^{-}_{t})$. 
Ties are broken using some deterministic rule. 

From Definition \ref{def:strong_consistency}, $S_{\chi',T}$ is upper bounded by 1. Therefore, 
it is sufficient to prove a lower bound on $S_{\chi',T}$ that converges to $1$ as $T \to \infty$. 
To this end, we rewrite $S_{\chi',T}$ as \cite{Yang02}
\begin{equation}
\label{eq:upperbound}
S_{\chi',T}=\frac{\sum_{t=1}^{T}f_{\hat{m}_{t}}(k_{t}^{-})}{\sum_{t=1}^{T}f_{m_{t}^{*}}(k_{t}^{-})}  
+\frac{\sum_{t=1}^{T}(f_{m_{t}}(k_{t}^{-})-f_{\hat{m}_{t}}(k_{t}^{-}))}{\sum_{t=1}^{T}f_{m_{t}^{*}}(k_{t}^{-})} \leq 1. %
\end{equation}
By Assumption \ref{as:expected_reward}, it follows from (\ref{eq:upperbound}) that
\begin{equation}
\label{eq:smaller}
S_{\chi',T}\geq \frac{\sum_{t=1}^{T}f_{\hat{m}_{t}}(k_{t}^{-})}{\sum_{t=1}^{T}f_{m_{t}^{*}}(k_{t}^{-})}-\frac{\frac{1}{T}\sum_{t=1}^{T}B\mathbb{I}_{\left \{m_{t}\neq \hat{m}_{t} \right \}}}{\frac{1}{T}\sum_{t=1}^{T}f_{m_{t}^{*}}(k_{t}^{-})}.%
\end{equation}
The remainder of the proof consists of two parts. In the first part we show that 
\begin{equation}
\label{eq:one}
\frac{\frac{1}{T}\sum_{t=1}^{T}B\mathbb{I}_{\left \{m_{t}\neq \hat{m}_{t} \right \}}}{\frac{1}{T}\sum_{t=1}^{T}f_{m_{t}^{*}}(k_{t}^{-})}\overset{\textup{a.s.}}{\rightarrow} 0,~\textup{as}~~T\rightarrow \infty, 
\end{equation}
while the second part deals with
\begin{equation}
\label{eq:two}
\frac{\sum_{t=1}^{T}f_{\hat{m}_{t}}(k_{t}^{-})}{\sum_{t=1}^{T}f_{m_{t}^{*}}(k_{t}^{-})}\overset{\textup{a.s.}}{\rightarrow} 1,~\textup{as}~~T\rightarrow \infty.
\end{equation}
Combining (\ref{eq:one}) and (\ref{eq:two}) with (\ref{eq:smaller}) and $S_{\chi',T} \leq 1$ 
proves the strong consistency.\\
(i) By Assumption \ref{as:expected_reward}, $\sum_{t=1}^{T}f_{m_{t}^{*}}(k_{t}^{-})$ is 
positive. As a result, $\frac{1}{T}\sum_{t=1}^{T}f_{m_{t}^{*}}(k_{t}^{-})$ converges 
to $\textup{E}_{t}\left \{f_{m_{t}^{*}}(k_{t}^{-})\right\}> 0$ almost surely. Hence, it suffices 
to show that $ \frac{1}{T}\sum_{t=1}^{T}B\mathbb{I}_{\left \{m_{t}\neq \hat{m}_{t}\right\}}\rightarrow 0$, 
almost surely. To show this, we consider the worst-case; that is, we assume that 
in all exploration trials, inferior arms are selected (i.e. the best-response is never 
selected by chance). Therefore
\begin{equation}
\label{eq:limit}
\lim_{T \to \infty}\frac{1}{T}\sum_{t=1}^{T}B\mathbb{I}_{\left \{m_{t}\neq \hat{m}_{t}\right\}}=\lim_{R \to \infty} \frac{\sum_{r=1}^{R}\left \lceil T'_{r}Z_{r} \right \rceil}{\sum_{r=1}^{R}T'_{r}} =0,
\end{equation}
where the second equality follows from Assumption \ref{as:bandit_game} and Lemma \ref{lm:Assumption}. 
This proves (\ref{eq:one}).\\
(ii) First, we note that (\ref{eq:two}) is equivalent to \cite{Yang02}\footnote{This part of the proof is almost 
identical to \cite{Yang02}; the difference is that here we use the fact that each action and also each joint 
action profile is played infinitely often (Lemma \ref{lm:numberPlay}) in order to complete the proof.}
\begin{equation}
\label{eq:equivalence}
\frac{\sum_{t=1}^{T}(f_{\hat{m}_{t}}(k_{t}^{-})-f_{m_{t}^{*}}(k_{t}^{-}))}{\sum_{t=1}^{T}f_{m_{t}^{*}}(k_{t}^{-})}\overset{\textup{a.s.}}{\rightarrow} 0,~\textup{as}~~T\rightarrow \infty.
\end{equation} 
Moreover, by (\ref{eq:upperbound}), (\ref{eq:smaller}) and (\ref{eq:one}), we conclude that
\begin{equation}
\label{eq:minus}
\frac{\sum_{t=1}^{T}(f_{\hat{m}_{t}}(k_{t}^{-})-f_{m_{t}^{*}}(k_{t}^{-}))}{\sum_{t=1}^{T}f_{m_{t}^{*}}(k_{t}^{-})}\leq 0.
\end{equation} 
Clearly,
\begin{equation}
\label{eq:decomposition}
\begin{aligned}
f_{\hat{m}_{t}}(k_{t}^{-})-f_{m_{t}^{*}}(k_{t}^{-})& = f_{\hat{m}_{t}}(k_{t}^{-})-\hat{f}_{\hat{m}_{t},t-1}(k_{t}^{-})+\hat{f}_{\hat{m}_{t},t-1}(k_{t}^{-})\\ 
&-\hat{f}_{{m_{t}^{*}},t-1}(k_{t}^{-})+\hat{f}_{m_{t}^{*},t-1}(k_{t}^{-})-f_{m_{t}^{*}}(k_{t}^{-}).\\
\end{aligned}
\end{equation} 
On the other hand, for every trial $t$, $\hat{f}_{\hat{m}_{t},t-1}(k_{t}^{-})\geq \hat{f}_{m^{*}_{t},t-1}(k_{t}^{-})$ 
holds. Hence we can write \cite{Yang02}
\begin{equation}
\label{eq:upBound}
\begin{aligned}
f_{\hat{m}_{t}}(k_{t}^{-})&-f_{m_{t}^{*}}(k_{t}^{-})\\
\geq f_{\hat{m}_{t}}& (k_{t}^{-})-\hat{f}_{\hat{m}_{t},t-1}(k_{t}^{-})
-f_{{m_{t}^{*}}} (k_{t}^{-})+\hat{f}_{{m_{t}^{*}},t-1}(k_{t}^{-})\\
\geq -2~&\sup_{1\leq m\leq M} \left \|\hat{f}_{m,t-1}(k_{t}^{-})-f_{m}(k_{t}^{-})\right \|_{\infty}.
\end{aligned}
\end{equation} 
This yields
\begin{equation}
\label{eq:upBoundTwo}
\begin{aligned}
& \frac{\sum_{t=1}^{T}(f_{\hat{m}_{t}}(k_{t}^{-})-f_{m_{t}^{*}}(k_{t}^{-}))}{\sum_{t=1}^{T}f_{m_{t}^{*}}(k_{t}^{-})}\\ 
&\geq \frac{\frac{-2}{T}\sum_{t=1}^{T}\sup_{1\leq m\leq M}\left \|\hat{f}_{m,t-1}(k_{t}^{-})-f_{m}(k_{t}^{-})\right\|_{\infty }}{\frac{1}{T}\sum_{t=1}^{T}f_{m_{t}^{*}}(k_{t}^{-})}.
\end{aligned}
\end{equation} 
For brevity, let us rewrite (\ref{eq:upBoundTwo}) in a shorter form as $a\geq b$. By Assumption \ref{as:consistent_reg},  $\|\hat{f}_{m,T}(k^{-})-f_{m}(k^{-})\|_{\infty}\to 0$ as $T\to \infty$. 
However, in order to use 
this assumption, we need to ensure that not only each arm, but also each joint action profile is 
played infinitely many times, as $T \to \infty$. This is established in Lemma \ref{lm:numberPlay}. 
Therefore, the right-hand side of (\ref{eq:upBoundTwo}) converges to zero, i.e. $b \to 0$ and hence 
$a \geq 0$. On the other hand, by (\ref{eq:minus}), the left-hand side is upper-bounded by zero, that 
is $a \leq 0$. As a result, (\ref{eq:equivalence}) follows, which completes the second part of the proof.


%
\subsection{Proof of Theorem \ref{th:Convergence}}\label{subsec:th_convergence}
Consider a K-player MAB game, as described in Section \ref{subsec:MP_MAB}. By Theorem 
\ref{th:CorrConvergence}, if each player plays by best responding to a calibrated 
forecast of the joint action profile of opponents, then 
\begin{equation}
\inf_{\pi \in \mathfrak{C}}\sum_{\textbf{m}}\left |\hat{\pi}_{T}(\textbf{m}) -\pi(\textbf{m})\right | \to 0, 
\end{equation}
as $T \to \infty$. We refer to this selection strategy as $\chi'$. 
In order to prove the Theorem, we show that our strategy $\chi$, in which 
the true expected rewards of joint action profiles are not known and are gradually learned 
by exploration, exhibits the same convergence characteristics as $\chi'$. 

First, we re-arrange the K-player MAB game to a two-agent game where the first agent is 
any player $k$ and the second agent is the set of its opponents, i.e. the set of $K-1$ 
players. For this game, any joint action profile of the two agents can be written as 
$(m,\textbf{m}^{-})$, where $m \in \left \{ 1,...,M \right \}$ and 
$\textbf{m}^{-} \in \bigotimes_{k=1}^{K-1}\left \{1,...,M \right \}$. 
Let $\hat{\pi}_{T}(m,\textbf{m}^{-})$ denote the fraction of time until $T$ in which some 
joint action $(m,\textbf{m}^{-})$ is played. According to selection strategy $\chi$, 
$\hat{\pi}_{T}(m,\textbf{m}^{-})$ can be written as 
\begin{equation}
\label{eq:farction}
\hat{\pi}_{T}(m,\textbf{m}^{-})=\hat{\pi}_{T,r}(m,\textbf{m}^{-})+\hat{\pi}_{T,i}(m,\textbf{m}^{-}),
\end{equation}
where $\hat{\pi }_{T,r}(m,\textbf{m}^{-})$ and $\hat{\pi }_{T,i}(m,\textbf{m}^{-})$ denote the fractions 
of time in which $(m,\textbf{m}^{-})$ is played by exploration (i.e. by chance), and by exploitation 
(i.e. according to the best response rule given by (\ref{eq:BestResponse})), respectively. 
According to Algorithm \ref{alg:Selection}, the total number of exploration trials is given 
by $\sum_{r=1}^{R}\left \lceil T'_{r}Z_{r} \right \rceil$. Moreover, by Assumption \ref{as:bandit_game}, 
we know 
\begin{equation}
\lim_{R \to \infty} \frac{\sum_{r=1}^{R}\left \lceil T'_{r}Z_{r} \right \rceil}{\sum_{r=1}^{R}T'_{r}}=0.
\end{equation}
This implies that
\begin{equation}
\label{eq:farctionTwo}
\hat{\pi}_{T,r}(m,\textbf{m}^{-})=0,
\end{equation}
for $T \to \infty$. Therefore, in the limit, $\hat{\pi}_{T,r}(m,\textbf{m}^{-})$ can be neglected when calculating 
the empirical frequencies of plays, and
\begin{equation}
\label{eq:farctionThree}
\hat{\pi}_{T}(m,\textbf{m}^{-})=\hat{\pi }_{T,i}(m,\textbf{m}^{-})
\end{equation}
holds asymptotically. \\
In order to complete the proof, it is sufficient to show that after some finite time, the 
actions taken by the player based on $\hat{f}_{m,k,t}$ are equal to those based on $f_{m,k}$. 
By Assumption \ref{as:expected_reward}, we know that, with probability 1, there exists an 
$\nu>0$ so that for every $m \in \left \{1,...,M \right \}$ and after a time point $\theta<\infty$, 
\begin{equation}
\label{eq:Conv-Lemma}
\|\hat{f}_{m,k}(k^{-})-f_{m,k}(k^{-})\|<\nu,
\end{equation}
holds for all $t>\theta$. At the same time, according to our system model and by 
Assumption \ref{as:expected_reward}, the reward functions are bounded, and the 
action space and memory are finite. This implies that if (\ref{eq:Conv-Lemma}) 
holds, then the actions of the player evolve as if it were aware of the 
true expected reward of each joint action profile, which completes the proof.

%

\bibliographystyle{IEEEbib}
\bibliography{Main}
\end{document}